\documentclass[11pt]{article}
\usepackage{epsfig}
\usepackage{graphicx,amssymb,amsmath}
\usepackage{latexsym}
\usepackage[dvips]{color}
\begin{document}
\newcommand{\M}{\mbox{m}}
\newcommand{\n}{\mbox{$n_f$}}
\newcommand{\EP}{\mbox{$e^+$}}
\newcommand{\EM}{\mbox{$e^-$}}
\newcommand{\EPEM}{\mbox{$e^+e^{-}$}}
\newcommand{\EMEM}{\mbox{$e^-e^-$}}
\newcommand{\GG}{\mbox{$\gamma\gamma$}}
\newcommand{\GE}{\mbox{$\gamma e$}}
\newcommand{\GP}{\mbox{$\gamma e^+$}}
\newcommand{\TEV}{\mbox{TeV}}
\newcommand{\GEV}{\mbox{GeV}}
\newcommand{\LGG}{\mbox{$L_{\gamma\gamma}$}}
\newcommand{\LGE}{\mbox{$L_{\gamma e}$}}
\newcommand{\LEE}{\mbox{$L_{ee}$}}
\newcommand{\LEPEM}{\mbox{$L_{e^+e^-}$}}
\newcommand{\WGG}{\mbox{$W_{\gamma\gamma}$}}
\newcommand{\WGE}{\mbox{$W_{\gamma e}$}}
\newcommand{\EV}{\mbox{eV}}
\newcommand{\CM}{\mbox{cm}}
\newcommand{\MM}{\mbox{mm}}
\newcommand{\NM}{\mbox{nm}}
\newcommand{\MKM}{\mbox{$\mu$m}}
\newcommand{\SEC}{\mbox{s}}
\newcommand{\CMS}{\mbox{cm$^{-2}$s$^{-1}$}}
\newcommand{\MRAD}{\mbox{mrad}}
\newcommand{\IND}{\hspace*{\parindent}}
\newcommand{\E}{\mbox{$\epsilon$}}
\newcommand{\EN}{\mbox{$\epsilon_n$}}
\newcommand{\EI}{\mbox{$\epsilon_i$}}
\newcommand{\ENI}{\mbox{$\epsilon_{ni}$}}
\newcommand{\ENX}{\mbox{$\epsilon_{nx}$}}
\newcommand{\ENY}{\mbox{$\epsilon_{ny}$}}
\newcommand{\EX}{\mbox{$\epsilon_x$}}
\newcommand{\EY}{\mbox{$\epsilon_y$}}
\newcommand{\BI}{\mbox{$\beta_i$}}
\newcommand{\BX}{\mbox{$\beta_x$}}
\newcommand{\BY}{\mbox{$\beta_y$}}
\newcommand{\SX}{\mbox{$\sigma_x$}}
\newcommand{\SY}{\mbox{$\sigma_y$}}
\newcommand{\SZ}{\mbox{$\sigma_z$}}
\newcommand{\SI}{\mbox{$\sigma_i$}}
\newcommand{\SIP}{\mbox{$\sigma_i^{\prime}$}}
\newcommand{\be}{\begin{equation}}
\newcommand{\ee}{\end{equation}}
\newcommand{\bc}{\begin{center}}
\newcommand{\ec}{\end{center}}
\newcommand{\bi}{\begin{itemize}}
\newcommand{\ei}{\end{itemize}}
\newcommand{\ben}{\begin{enumerate}}
\newcommand{\een}{\end{enumerate}}
\newcommand{\bm}{\boldmath}

\title{\bf The Photon collider at ILC: \\ status, parameters and technical
  problems~\thanks{Presented at   PLC2005, Kazimierz, Poland, September 2005}}
\author{V.I.~Telnov~\thanks{telnov@inp.nsk.su} \\[2mm] 
{\it Budker Institute of Nuclear Physics, 630090 Novosibirsk, Russia} }
\date{}
\maketitle
\begin{abstract}
  This paper is the second part of my overview on photon colliders
  given at the conference ``The photon: its first hundred years and the future''
  (PHOTON2005 + PLC2005). The first paper~\cite{telnov-ph05} describes
  the first 25 years of the history and evolution of photon colliders. The
  present paper considers the photon collider at the ILC:
  possible parameters, technical problems and present status.
\end{abstract}

\section{Introduction}

There is a consensus in the particle-physics community that the next large
project after the LHC should be a linear \EPEM\ collider.  Due to the high
cost of such a collider it has been agreed to build a single collider at the energy
$2E_0=0.5$--$1$ TeV instead of the three regionally developed colliders,
TESLA, NLC and JLC. In 2004, the International Linear Collider (ILC),
based on the superconducting TESLA-like technology, was inaugurated.
The project will be approved for construction after observation of interesting
physics in this energy region by the LHC, which starts operation in 2007.
At present, the development of the ILC and its detectors is proceeding under
the guidance of the ILCSC, GDE and WWS committees. The next steps are: the choice
of the baseline configuration, the reference design, site selection,
and the conceptual and technical designs.

\begin{figure}[htb]
\vspace{0.cm}
\centering
\includegraphics[width=8cm]{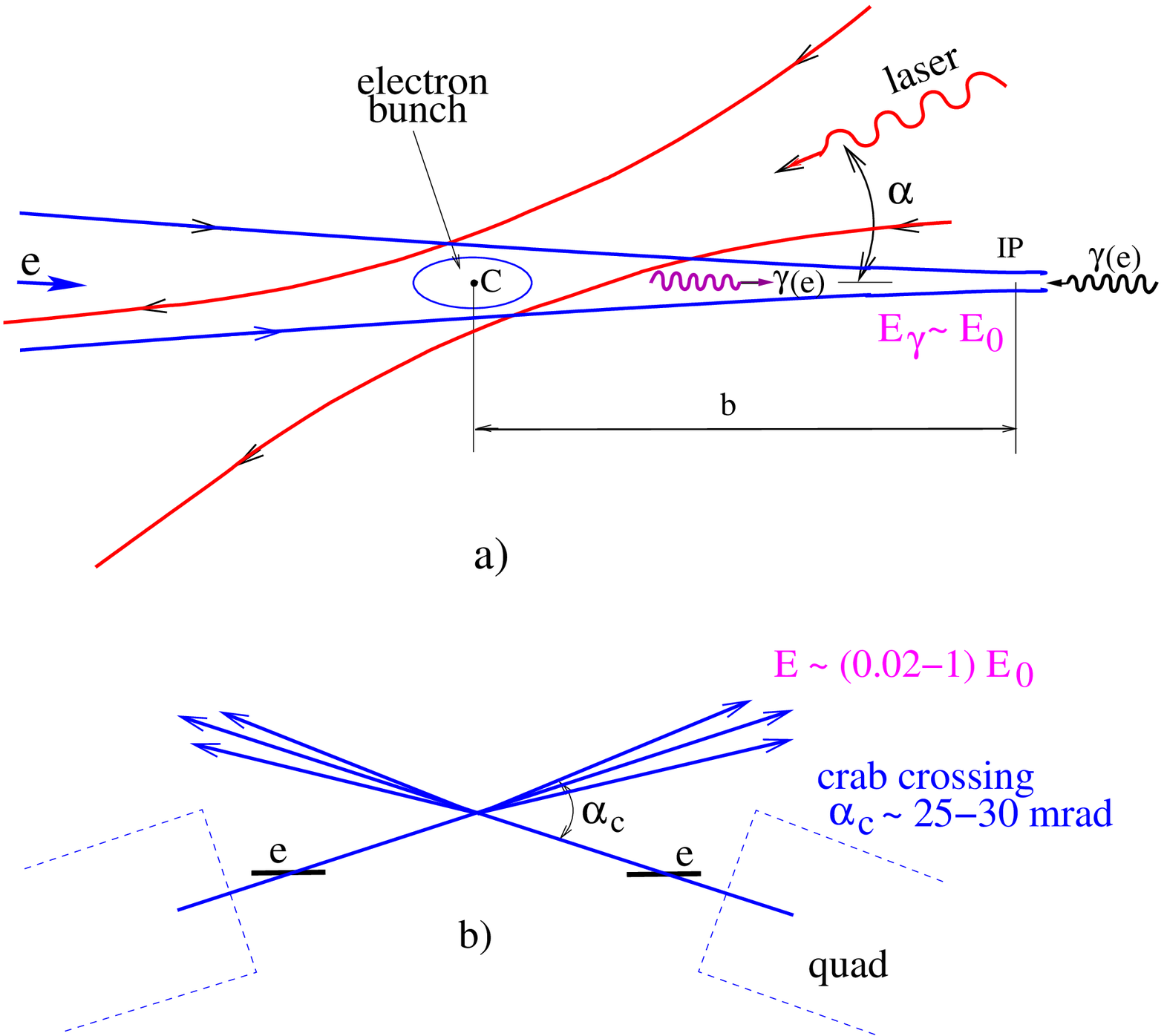}
\caption{Scheme of \GG,\GE\ collider}
\vspace{-0.4cm}
\label{scheme}
\end{figure}

It is well understood that in addition to \EPEM\ physics, linear
colliders provide a unique opportunity to study \GG\ and \GE\ 
interactions at high energy and
luminosity~\cite{GKST81,GKST83,TESLATDR}.  High-energy photons are
obtained by "conversion" of electrons into high-energy photons using
Compton scattering of laser light at a small distance from the
interaction point (IP), Fig.~\ref{scheme}. The photon collider is
highly appreciated by the physics community: more than 20\% of all
publications on linear colliders are devoted to photon colliders (in
spite of the fact that at present this activity is not funded). 

The motivation for the photon collider is very strong:\\[-7mm]
\bi
\item The physics is very  rich [4--11]\,: \\[-8mm]
  \bi
\item the energy is lower than in \EPEM\ collisions only
by 10--20\%; \\[-5mm]
\item the number of interesting events is similar or even greater; \\[-5mm]
\item access to higher particle masses (single resonances in $H$, $A$,
etc., in \GG, heavy charged and light neutral (SUSY, etc.) in
  \GE);  \\[-5mm]
\item in some scenarios, heavy $H/A$-bosons will be seen only in
\GG;\\[-5mm]
\item higher precisions for some important phenomena; \\[-5mm]
\item different (from \EPEM) types of reactions;  \\[-5mm]
\item highly polarized photons; \\[-5mm] \ei

\item there are no technical stoppers; the risk is small because in
  the case of technical problems the detector can continue taking data in the
  \EPEM\ mode of collisions; the relative incremental cost is small;\\[-6mm]

\item there is a great interest in the physics community to such experiments.
\ei

It is assumed that during the first several years of ILC operation 
all ILC detectors (whether one or
two) will run in \EPEM\ mode. Then, one of the interaction regions
(IRs) and detectors will be modified for operation in the \GG, \GE\ mode.
In other words, \EPEM\ collisions are considered ``baseline,'' while \GG,
\GE\ is seen as an ``option''. Now, ``option'' is quite a misleading term, one that
some people (including some of the ILC leaders) understand as an absolute
priority of \EPEM\ in all decisions and considerations, while the photon collider
is seen as being far in the future and thus not requiring any attention at this time. 
Moreover, in order
to reduce the ILC cost, there is a tendency to simplify the ILC design to a bare minimum:
one IP, one detector, no options. One physicist's comment regarding this was, ``we do not
need such a bicycle!''

Yes, the photon collider is part of the second stage of the ILC, but it has
many specific features (see the list below) that strongly influence
the baseline ILC configuration and the parameters of practically all
of its subsystems. These requirements should be included into the baseline
project from the very beginning---otherwise the upgrade from \EPEM\
to \GG, \GE\ will be very costly (or even impossible at all) and
(or) the parameters (such as the luminosity) of the photon collider 
will be much worse than in
the case of a properly optimized design. All this means that {\bf the 
photon collider should be considered from the beginning as an integral 
part of the ILC project!}

As for the cost, it is hard to imagine a project as cost-effective
as the ILC photon collider. It practically doubles the ILC physics program
while increasing the total cost project only by ${\cal O }(3\%)$. It is my firm belief
that it will be no problem at all to convince the funding agencies that such an 
small increment of
the ILC cost, which allows the ILC to study new phenomena in new types of
collisions, is extremely well justified. 

The next few years before the completion of the final ILC technical design 
are very important for the photon collider. All machine features required for the photon
collider should be properly included in the basic ILC design.  Of
course, it is also important to continue the development of the
physics program and to start, at last, the development of
the laser system, which is a key element of the photon collider---but
even more urgent are the accelerator and interaction-region aspects that
influence the design of the entire ILC project.

The most comprehensive description of the photon collider available at
present is part of the TESLA TDR \cite{TESLATDR}; almost all
considerations done for TESLA are valid for the ILC as well.  In the
following sections I consider the most important problems of the photon collider
that need special and careful attention of ILC designers.

\section{Requirements for the ILC design}
The photon collider imposes several special requirements that should
be taken into account in the baseline ILC design:
\bi

\item For the removal of  disrupted beams, the crab-crossing angle at one of
  the interaction regions should be about 25 mrad; the ILC configuration
  should allow an easy transition between \EPEM\ and \GG\
  modes of operation;

\item The \GG\ luminosity is nearly proportional to the geometric
  \EMEM\ luminosity, so the product of the horizontal and vertical
  emittances should be as small as possible (this translates into
  requirements on the damping rings and beam-transport lines);

\item The final-focus system should provide a beam-spot size at the
  interaction point that is as small as possible (compared to the \EPEM\ case, the
  horizontal $\beta$-function should be smaller by one order of
  magnitude);

\item The very wide disrupted beams should be transported to the beam
  dumps with acceptable losses. The beam dump should be able to
  withstand absorption of a very narrow photon beam after the Compton
  scattering;

\item  The detector design should allow easy replacement of elements in the
forward region ($<$100 mrad);

\item Space for the laser and laser beam lines has to be reserved.
\ei

Ignorance of any of these requirements can result in the future in
a significant increase of the cost, in loss of time and in poor 
photon-collider parameters.
\section{Photon collider luminosity}
There are three luminosity problems at photon colliders: 1) obtaining
high luminosities, 2) stabilization of beam collisions, 3) measurement
of the luminosities, all these problems are discussed below. The most
important and urgent at this time is the first problem.
\subsection{Towards high \GG, \GG\ luminosities}
The \GG\ luminosity at the ILC energies is determined by the geometric
luminosity of electron beams~\cite{Tfrei,TEL2001,TESLATDR}.
There is an approximate general rule: the luminosity in the high-energy
part of the spectrum $\LGG \sim 0.1 L_{ \rm geom}$, where $ L_{ \rm
  geom}=N^2 \nu \gamma / 4\pi \sqrt{\ENX \ENY\ \beta_x \beta_y}$. So,
to maximize the luminosity,
one needs the smallest beam emittances \ENX, \ENY\ and beta-functions at
the IP, approaching the bunch length. Compared to the \EPEM\ case,
where the minimum transverse beam sizes are determined by
beamstrahlung and beam instability, the photon collider needs
a smaller product of horizontal and vertical emittances and a smaller
horizontal beta-function.

The ``nominal'' ILC beam parameters are: $N= 2\times 10^{10}$,
$\sigma_z=0.3$ mm, $\nu= 14100$ Hz, $\ENX = 10^{-5}$ m, $\ENY = 4
\times 10^{-8}$ m.  Obtaining $\beta_y \sim \sigma_z=0.3$ mm is not a
problem, while the minimum value of the horizontal $\beta$-function
is restricted by chromo-geometric aberrations in the final-focus
system~\cite{TESLATDR}. For the above emittances, the limit on the
effective horizontal $\beta$-function is about 5
mm~\cite{TEL-Snow2005,Seryi-snow}.  The expected \GG\ luminosity
$\LGG(z> 0.8z_m) \sim 3.5 \times 10^{33}$ \CMS\ $\sim 0.17\,\LEPEM$
(here the nominal $\LEPEM = 2\times 10^{34}$\CMS)~\cite{TEL-Snow2005}.
Taking into account the fact that many cross sections in \GG\ are larger than those
in \EPEM\ collisions by one order of magnitude, the event rate
will be somewhat larger than in \EPEM\ collisions.

The above-mentioned luminosity corresponds to the beam parameters optimized for
the \EPEM\ collisions where the luminosity is determined by collision
effects. The photon collider has no such restrictions and can work
with much smaller beam sizes. The horizontal beam size at the
parameters under consideration is $\sigma_x \approx 300$ nm, while the
simulation shows that the photon collider at such energies can work
even with $\sigma_x \sim 10$ nm  without fundamental
limitations~\cite{TESLATDR}. So, we are very far from the physical
limit and should do all that is possible to minimize transverse beam sizes
at the photon collider!

   It should be noted that the minimum $\beta_x$ depends on the
   horizontal emittance. It is about 5 mm for the nominal
   emittance and 3.7 (2.2) mm for emittances reduced by a factor of 2 (4), 
   respectively. In the TESLA project, we considered  emittances
   close to the latter case: $\ENX=0.25\times 10^{-5}, \ENY=3\times
   10^{-8}$ m, which gives a \GG\ luminosity that is a factor of 3.5 higher!
   
   The beams are produced in the damping rings (DR), so the minimum emittances are
   determined by various physics effects such as quantum fluctuations
   in synchrotron radiation and intra-beam scattering (IBS). The
   latter is the most difficult to overcome. Where is the limit?  One of
   the possible way for reducing emittances is to decrease the damping
   time by adding wigglers~\cite{Wolski-snow}, which has not yet been
   considered in detail by experts. 

Damping rings are complex devices, so one should trust only careful studies. 
Nevertheless, I would like to make some rough estimates.

   The equilibrium emittance in the wiggler-dominated regime due to
   quantum fluctuations~\cite{Wiedemann}
\be \ENX \sim 3.3 \times 10^{-11}
   B_0^3(\mbox{T}) \lambda^2(\mbox{cm}) \beta_x (\mbox{m}) \; \mbox{m}.
\ee
   The damping time
\be \tau_d = \frac{3m^2c^3}{r_e^2 E B_0^2} =
   \frac{5.2 \times 10^{-3}}{E(\GEV)B_0^2(\mbox{T})}\; \mbox{sec}.
\ee If
   wigglers fill 1/3 of the DR then for $B_0=2$ T and $E=5$ GeV one
   gets $\tau=7.5\times10^{-4}$ sec, which is more than 20 times smaller
   than the damping time in existing designs.
   For $\lambda=10$ cm and $\beta_x=5$ m, the equilibrium normalized
   emittance due to synchrotron radiation is $\ENX=1.3\times 10^{-7}$
   m, which is 60 times smaller than the present nominal emittance. The
   vertical emittance will be much smaller as well. This does not take into account the IBS.

  So, there appears to be a lot of room for decreasing the damping
  time and thus decreasing emittances in $x$ and $y$, as well as
  $\beta_x$. Until $\beta_x$ and $\sigma_y$ are larger than their
  limits ($\sigma_z$ and 1 nm, respectively), there is a strong
  dependence of the luminosity on emittances ($L \propto 1/\sqrt{\ENX \ENY
\beta_x}$). The increase of the luminosity by a factor of 10 is not
  impossible with appropriate modifications to the damping rings! This would require
  more  RF peak power, but that problem is solvable. The turn shift
  due to the beam space charge may be unimportant due to strong
  damping. This looks promising and needs a serious consideration by DR
  experts\,!

  Let us assume a reduction (compared to the nominal beam
  parameters) of \ENX\ by a factor of 6, \ENY\ by a factor of 4 and
  $\beta_x$ down to 1.7 mm (it is possible for such emittances). Then,
  one can have the following parameters of the photon collider:
  $N=2\times 10^{10}$, $\nu= 14$ kHz, $\ENX=1.5\times 10^{-6}$ m,
  $\ENY= 1.\times 10^{-8}$ m, $\beta_x=1.7$ mm, $\beta_y=0.3$ mm, the
  distance between interaction and conversion regions is 1 mm,
  $\sigma_x=72$ nm, $\sigma_y=2.5$ nm, $L_{\rm geom}=2.5\times
  10^{35}$, $\LGG(z>0.8z_m) \sim 2.5\times 10^{34}$ \CMS\ $\sim 1.25
  L_{e^+e^-,\, nomin.}$.

  The resulting \GG\ luminosity is greater than that at the ``nominal'' ILC beam
  parameters by a factor of 8.5.  The statistics in \GG\ collisions would then
  be higher than in \EPEM\ by  one order of magnitude, which would
  open new possibilities  such as the study of Higgs
  self-coupling in \GG\ collisions just above the $\GG\ \to hh$
  threshold ~\cite{Jikia,Belusevic}.

\begin{figure}[hbt]
\centering
\vspace*{-0.6 cm}
\hspace{-0.5cm}   \epsfig{file=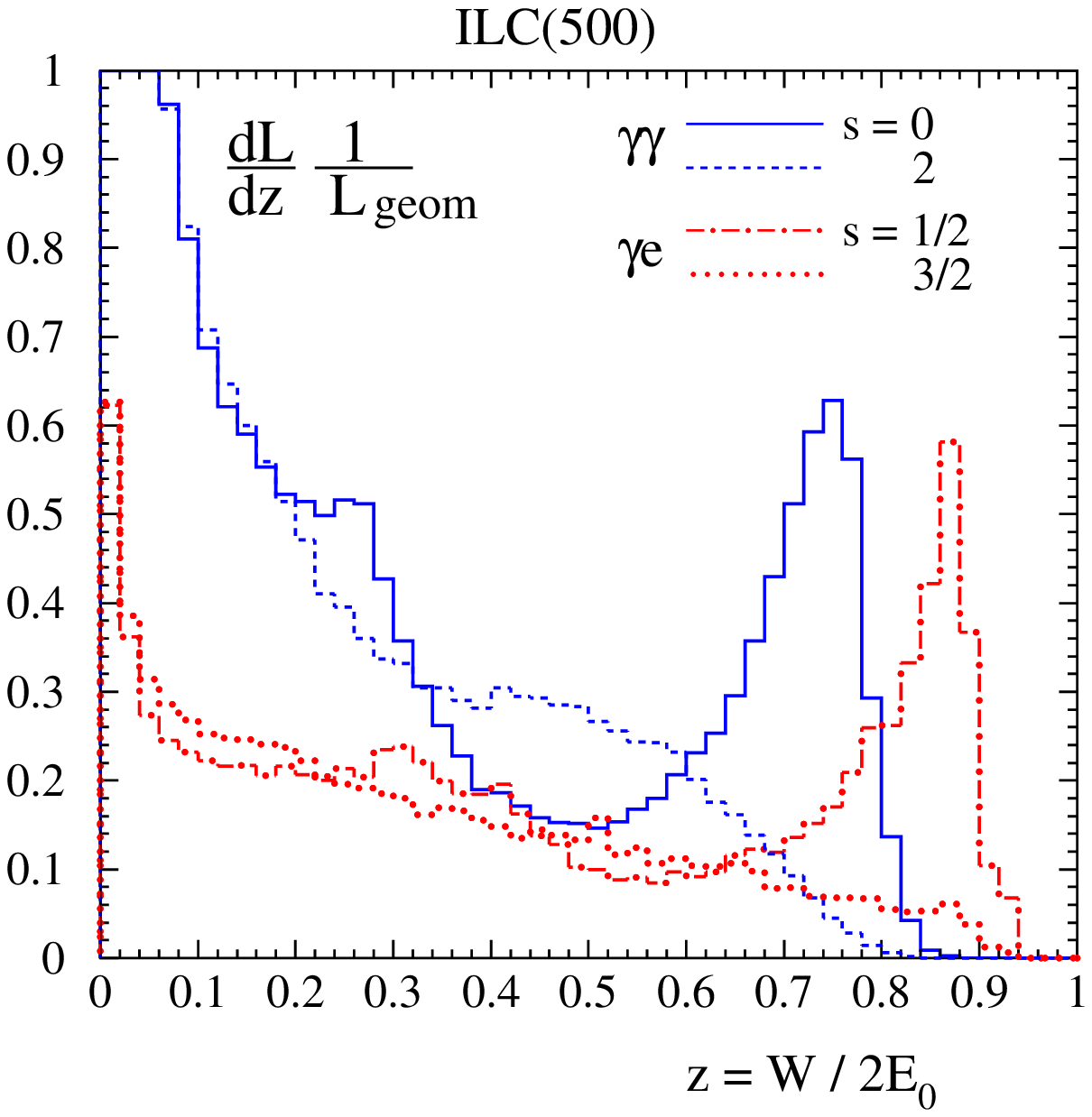,width=7.5cm,angle=0}
\hspace{-1cm} \epsfig{file=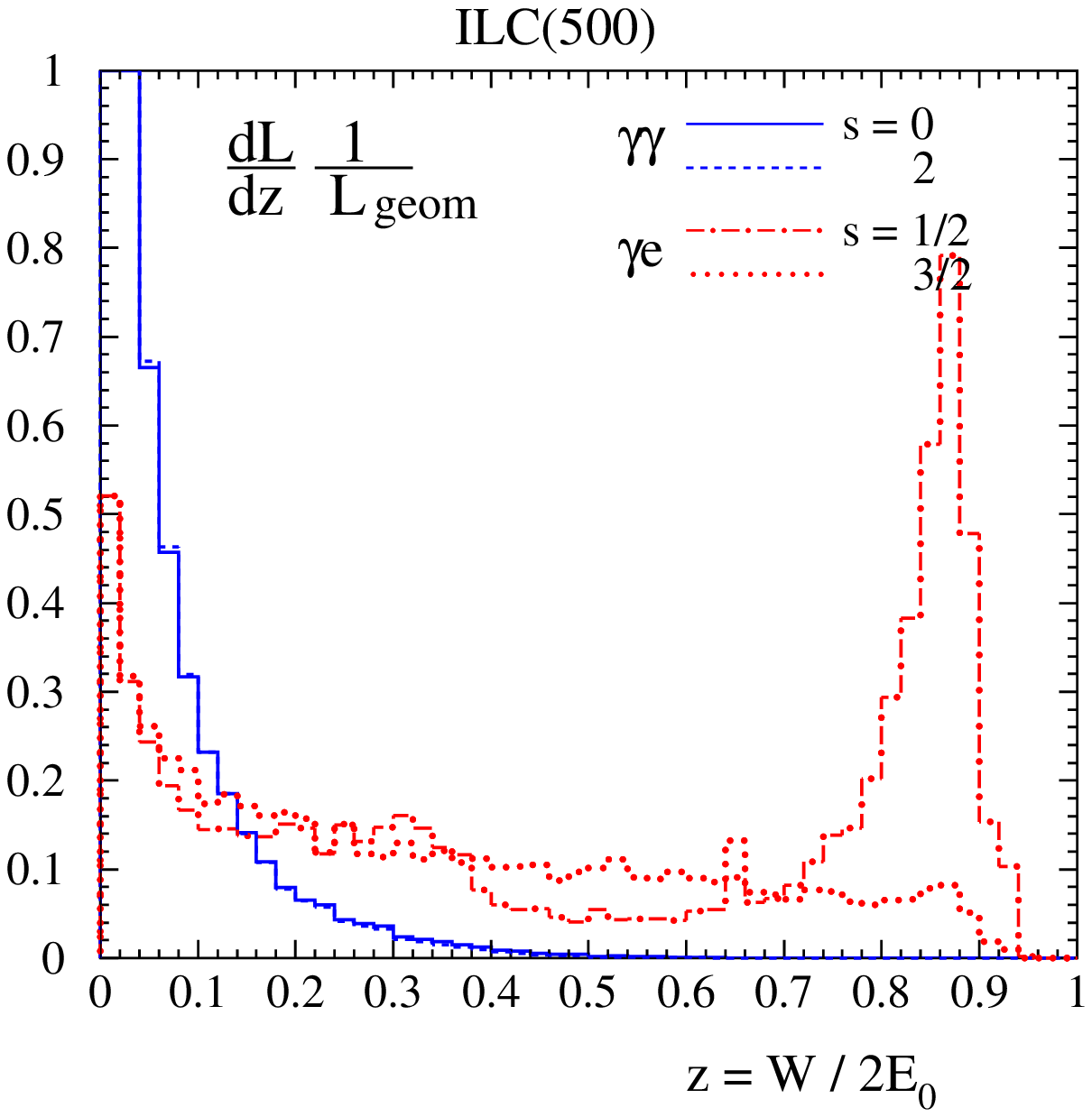,width=7.5cm,angle=0}
\vspace{-0.7cm}
\caption{ \GG, \GE\ luminosity spectra, left: both beams are converted
  to photons; right: only one beam is converted to photons. See
  parameters in the text. \GG.}
 \vspace{-0.2cm}
\label{luminosity}
\end{figure}

  Figures~\ref{luminosity} show simulated luminosity spectra for
  these parameters. All important effects are taken into account,
  including the increase of the vertical beam size in the detector
  field due to the crab crossing (Sect.~3).  In the figure on the right,
  only one of the electron beams is converted to photons, it is more
  preferable for \GE\ studies due to easier luminosity
  measurement~\cite{Pak} and smaller backgrounds. The
  corresponding luminosity $\LGE(z>0.8z_m)\sim 2.\times 10^{34}$ \CMS.
By increasing the distance between the conversion and interaction
regions, one can obtain a rather  monochromatic luminosity spectrum of
a reduced luminosity  for the study of QCD processes~\cite{TEL-mont1}.

I would like to stress once again that the parameters of the ILC damping rings are dictated
not by \EPEM, but by \GG\ collisions and a decision on the DR design
should be based on the dependence $\LGG = f(\mbox{DR cost})$. It could be that the
increase of the \GG\ luminosity by a factor of 8.5 as suggested above is too
difficult, but even x2 -- x3 improvement would be quite useful. This
is very important and urgent question\,!
\vspace{-0mm}

\subsection{Luminosity stabilization \label{stab}}
Beam collisions (luminosity) at linear colliders can be adjusted by
a feedback system that measures the beam-beam deflection using beam
position monitors (BPM) and corrects beam positions by fast kickers.
This method is considered  for \EPEM\ collisions and
is assumed for \GG\ as well~\cite{TESLATDR}.

\begin{figure}[hbt]
\vspace{-3mm}
  \centering
\includegraphics[height=4.5cm]{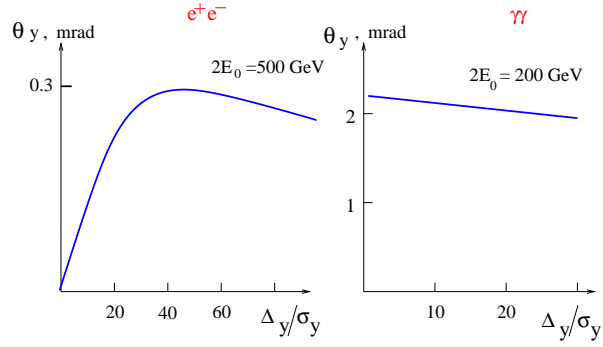}
\vspace{-3mm}
\caption{Typical beam-beam deflection in \EPEM\ and \GG\ collisions}
\label{f:defl1}
\end{figure}

\begin{figure}[hbt]
\centering \vspace{-2mm}
\includegraphics[width=8.5cm]{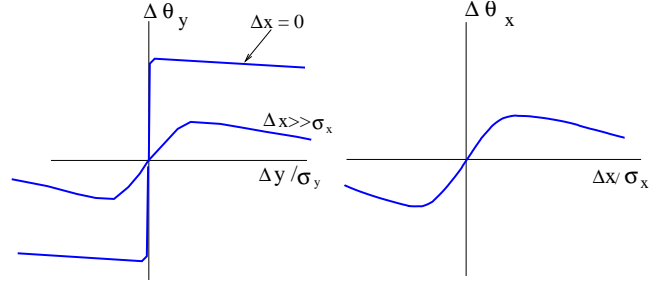}
\caption{beam-beam deflections in \GG\ collisions.}
\label{f:defl2}
\end{figure}

\begin{figure}[!htb]
\begin{minipage}{0.5\linewidth}
\hspace{0.3cm}\includegraphics[width=7cm,angle=0]{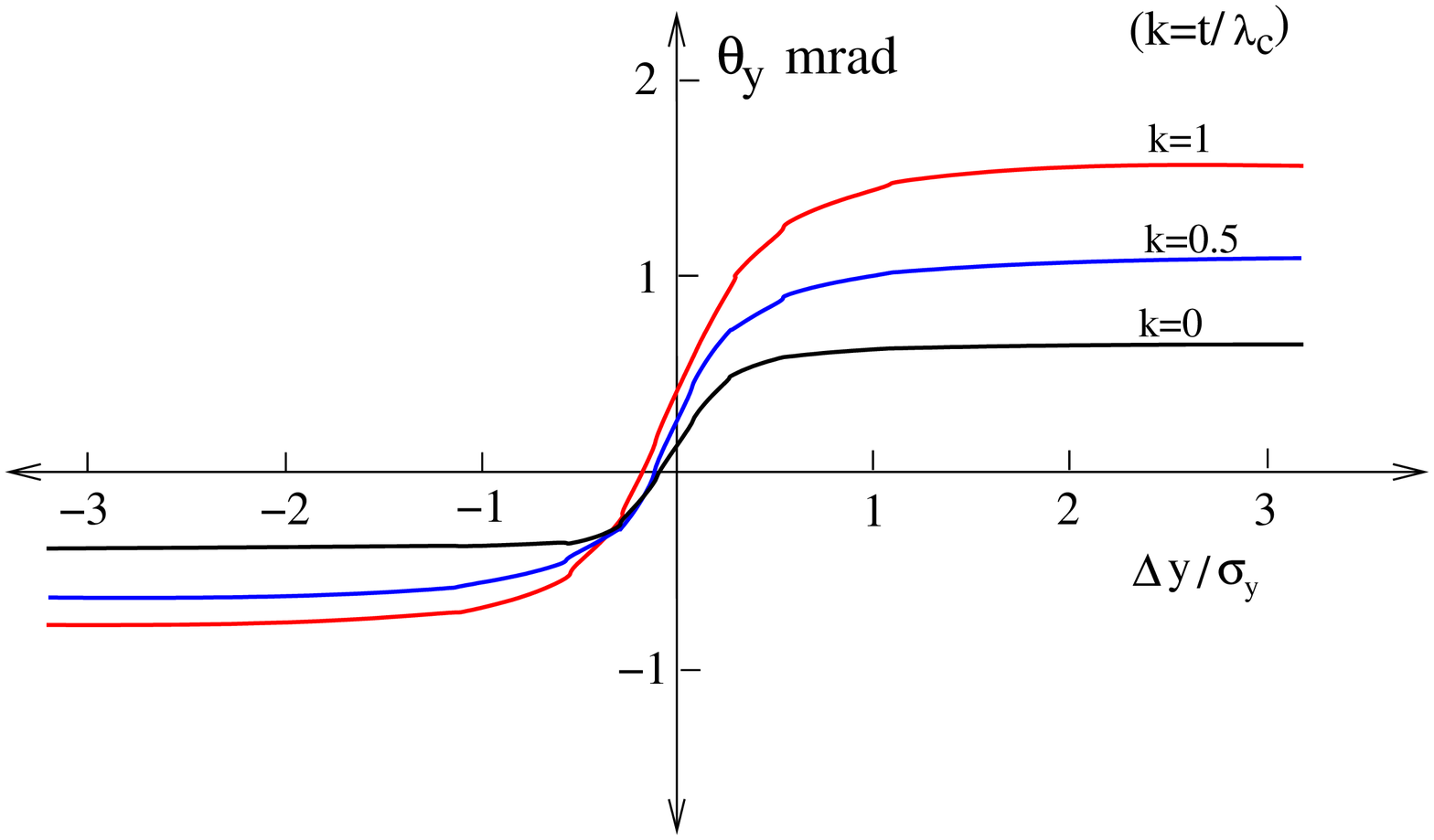}
\vspace{-0.6cm}
\caption{Deflection curves in \GG\ collisions for various conversion efficiencies.}
\label{f:defl3}
\end{minipage} \hspace{0.5cm}
\begin{minipage}{0.3\linewidth}
\vspace{-0.0cm}
\hspace{-0.cm}\includegraphics[width=5cm,angle=0]{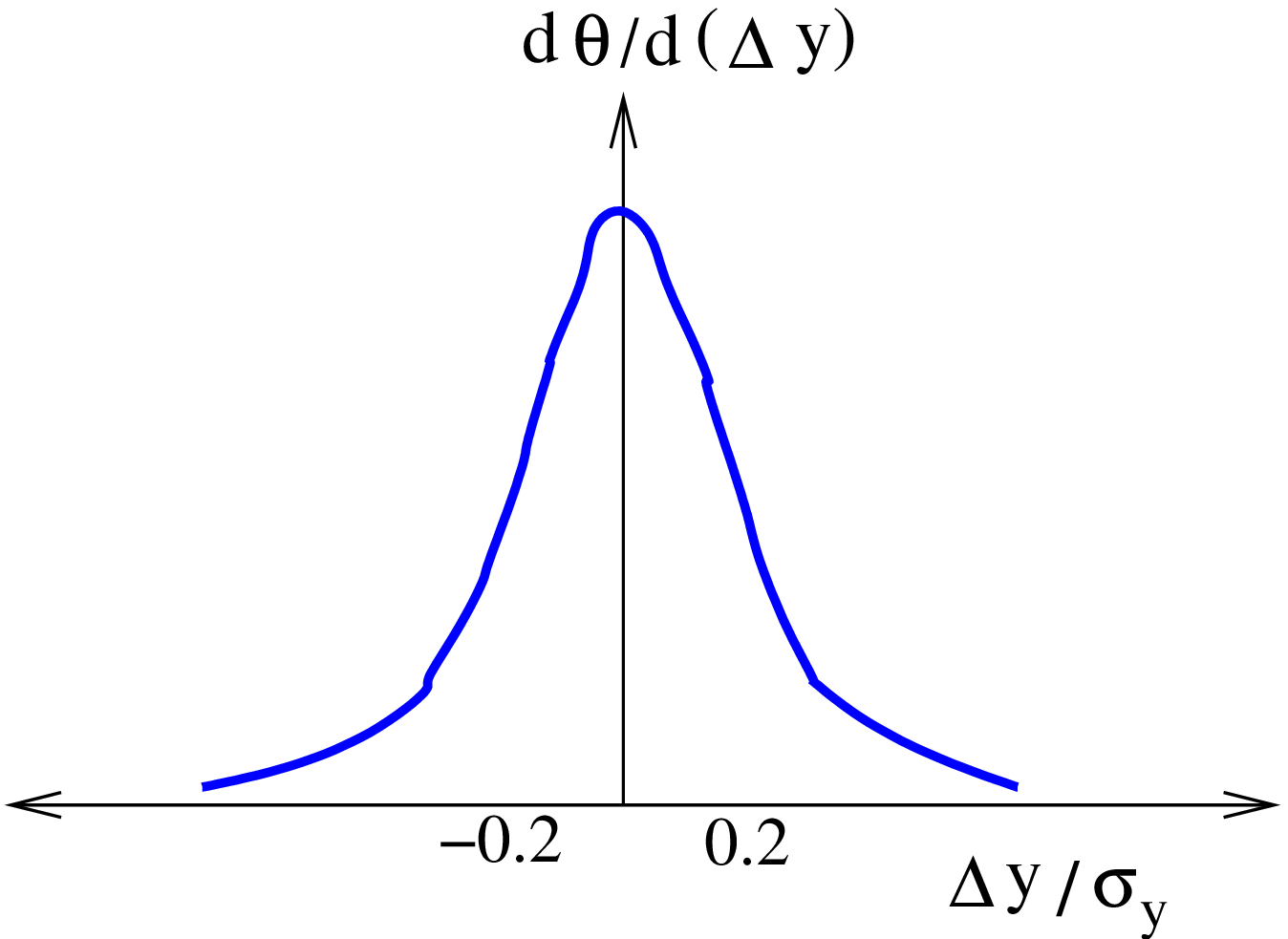}
\vspace{-0.6cm}
\caption{Derivative of the typical deflection curve}
\label{f:defl4}
\end{minipage}
\vspace{-0.cm}
\end{figure}

There are some differences between the \EPEM\ and \GG\ cases: \\[-0.7cm]
\begin{itemize}
\item In the \EPEM\ case, at small vertical displacements the beams
  attract each other and oscillate.  In the \GG\ case (\EMEM\ as well), the
  beams repel each other; as a result, the deflection angle is 
larger.\\[-0.6cm]
\item In the \GG\ case, due to Compton scattering, the average energy in the
  disrupted beam is several times smaller than the beam energy, which
  leads to a further increase of the deflection angle. \\[-0.7cm]
\item In  \GG\  collisions, $\sigma_x$ is several times smaller than in
 the \EPEM\ case. Due to a strong beam-beam instability,  the kick is large
and almost independent of the initial displacement.\\[-0.7cm]
\end{itemize}
There are two additional complications in \GG\ collisions: \\[-6mm]
  \begin{itemize}
\item The deflection curves depend on the conversion
    efficiency. \\[-7mm]
  \item Due to the crossing angle, the disrupted beam is deflected
    (mostly vertically) by the detector field. This additional deflection is
    comparable to the beam-beam deflection angle and also depends on
    the conversion probability, see Fig.~\ref{f:defl3}. This effect
    shifts the zero  point and creates a problem for 
stabilization of beam-beam collisions.\\[-7mm]
\end{itemize}
Typical deflection curves for \EPEM\ and \GG\ collisions are shown in
Fig.\ref{f:defl1}. For the \GG\ case, a smaller energy is taken in order to
emphasize the difference: the step-like behavior of the $\vartheta_y$
on the displacement $\Delta_y$. More general cases for \GG\ are shown
in Fig.\ref{f:defl2}.

We see that the deflection depends on the conversion coefficient,
the deflection curves are symmetric but shifted vertically due to the
detector field by some variable value.

How does one determine the vertical beam positions corresponding to the maximum
\GG\ luminosity? Here is the sloution. 
All deflection curves $\theta_y = f\,(\Delta y)$ have one common
feature: the derivatives $f^{\prime}$ reach the maximum at the point of
zero beam shifts where \LGG\ is maximum. This is illustrated in Fig.~\ref{f:defl4}:
the width of this "resonance" curve  is about $\pm 0.2 \sigma_y$
(for the cases being considered).

 The recipe for the \GG,\GE\ luminosity stabilization is the following:\\[-7mm]
\begin{enumerate}
\item Varying $\Delta y$ by decreasing steps (under software control), 
one finds the position of the jump in the deflection
curve with an accuracy of about $2\sigma_y$;\\[-7mm]
 \item Continue the scan with the step $0.1\sigma_y$ up and down
 around the point with the maximum derivative. The loss of \LGG\
 due to ``walking'' around the zero point will be negligibly small.\\[-7mm]
\item The horizontal zero point can be found in a similar way by
  varying the horizontal separation and measuring the horizontal deflection
  or by varying the horizontal separation and measuring the vertical
  deflection (the maximum vertical deflection corresponds to the zero
  horizontal separation).\\[-7mm]
\item In addition, the deflection in the detector field is very useful
  for optimization of the $e\to\gamma$ conversion. One just moves the
  laser beam and measures the vertical position of the outgoing beam in a 
  BPM at a distance of about 4 m from the IP. The maximum beam
  displacement corresponds to the maximum conversion efficiency.\\[-7mm]
\end{enumerate}

So, monitoring the the beam-beam deflection is a good method of stabilization of
the \GG, \GE\ luminosities at the ILC. The required algorithm does not appear to
be difficult to implement thanks to the large train length and
large inter-bunch spacing.

\subsection{Luminosity measurement}

The measurement of the luminosity at the photon collider is not an
easy task. The spectra are broad and one should measure the luminosity
and polarization as a function of energies $E_1, E_2$ of the colliding
particles~\cite{Pak}.  The luminosity spectrum and polarization can be
measured using various QED processes. These are  $\GG\to l^+l^-$
($l=e,\mu$)~\cite{TEL93,TESLATDR,Pak}, $\GG\to
l^+l^-\gamma$~\cite{Pak,Makarenko}  for \GG\ collisions and  $\GE\to\GE$ and
$\GE\to\;e^-\EPEM$ for \GE\ collisions~\cite{Pak}. Some other SM
processes can be useful as well.

There is one unsolved problem in the measurement of linear
polarizations in \GG\ collisions~\cite{TEL-mont2}. There exists an average
linear photon polarization at a given energy that can be measured
from the azimuthal distribution in $\GG\ \to l^+l^-$. However, in
addition to that, high energy photons have a linear polarization whose direction
depends on the photon scattering angle. The directions of linear
polarization of the colliding beams correlate, as the product
of the linear polarizations $l_{1,\gamma}l_{2,\gamma}$ (which is presented in
the cross section for the Higgs production) is quite large even for
completely unpolarized initial particles. This correlation is not
observable, neither in the cross section nor in the azimuthal
distribution of the process $\GG\ \to l^+l^-$. Fortunately, this
effect is absent at the high-energy peak of \GG\ luminosity, which can be
used for the Higgs study.

\section{The crossing angle for the photon collider}

\subsection{Minimum crossing angle}
The beam-crossing angle at the ILC is now one of the most hotly debated
issues. For experimentation with \EPEM\ beams, zero or small angles are
preferable, but in this case there are some problems with the removal
of used beams. At present, two IPs are considered for ILC, one with a small
crossing angle, 2 mrad, and the other with a large crossing angle, 14 or
20 mrad, where 14 is somewhat more preferred.

In \GG\ collisions, the outgoing beams are strongly disrupted and for
their removal a larger crossing angle is needed. In order to have better
compatibility with \EPEM\ this angle should be as small as possible.
So, there are several questions: \\[-7mm]
\bi
\item what is the minimum crossing angle suitable for \GG\ ? \\[-6.5mm]
\item is this angle compatible with \EPEM\ ?\\[-6.5mm]
\item what is the upgrade path from \EPEM\ to \GG\ ?\\[-6.5mm]
  \ei
  
  For removal of these disrupted beams one needs the crab-crossing
  angle to be larger than the disruption angle plus the angular size
  of the final quad, see Fig.~\ref{scheme}.  There is an additional
  requirement: the field outside the quad should be small in order to
  add a small deflection angle for the low-energy particles in the
  outgoing beam.

\begin{figure}[!htb]
\begin{minipage}{0.5\linewidth}
 \vspace{-0.9cm}
  \hspace{-0.cm}\includegraphics[width=7.5cm]{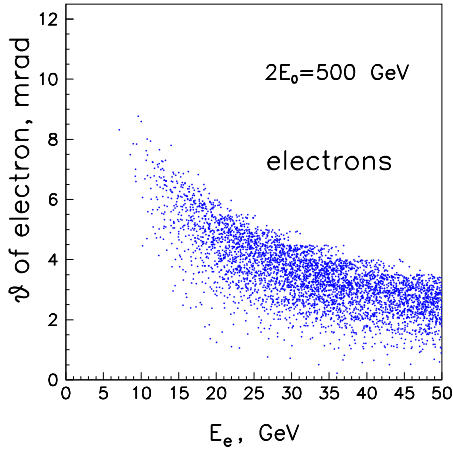}
 \vspace{-.8cm}
\end{minipage} \hspace{0.5cm}
\begin{minipage}{0.4\linewidth} \hspace{0mm}
 \vspace{-1.6cm}
\caption{Angles of disrupted electrons after Compton
    scattering and interaction with the opposing electron beam; $N=2\times
    10^{10}$, $\sigma_z=0.3$ mm.}  \hspace{0.5cm}
\label{e-angle}
\end{minipage}
\vspace{-0.cm}
\end{figure}

After passing the conversion and collision points, the electrons have
energy ranging from about 5 GeV up to $E_0$ and the horizontal disruption
angle up to about 10 mrad, see Fig.\ref{e-angle} (due to limited
statistics in simulation, about $10^5$ macroparticles, the maximum
angles should be multiplied by a factor of 1.2~ \cite{TESLATDR}).
Above this angle, the total energy of particles is less than that in
the secondary irremovable \EPEM\ background.  The disruption angle for
low-energy particles is proportional to $\sqrt{N/\sigma_z E}$~\cite{TEL90} and
depends very weakly on transverse beam sizes. 

\begin{figure}[!htb]
\vspace{-0.3cm}
\hspace{1.0cm} \includegraphics[width=6.7cm]{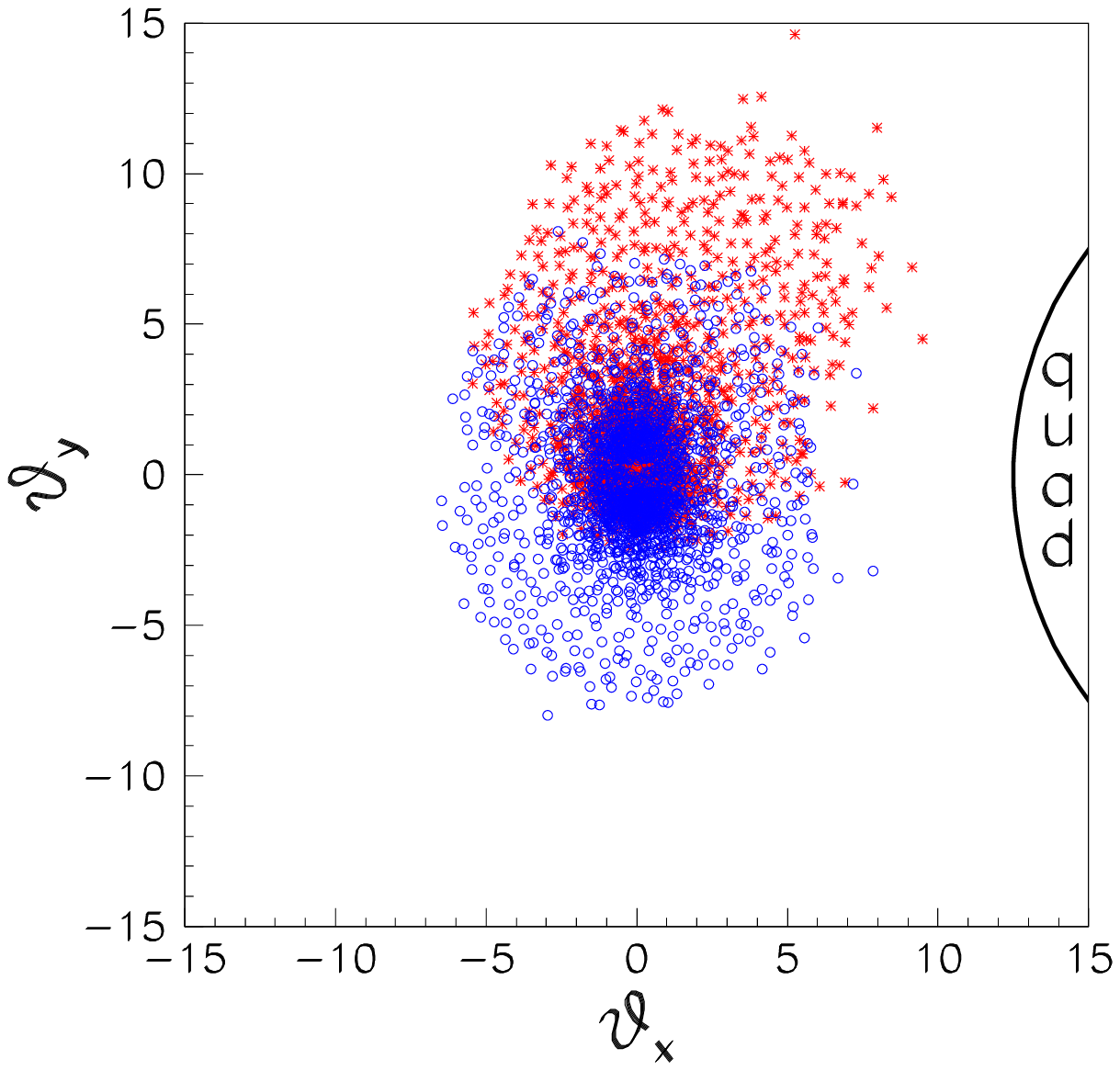}\hspace{-1cm}
\includegraphics[width=6.7cm]{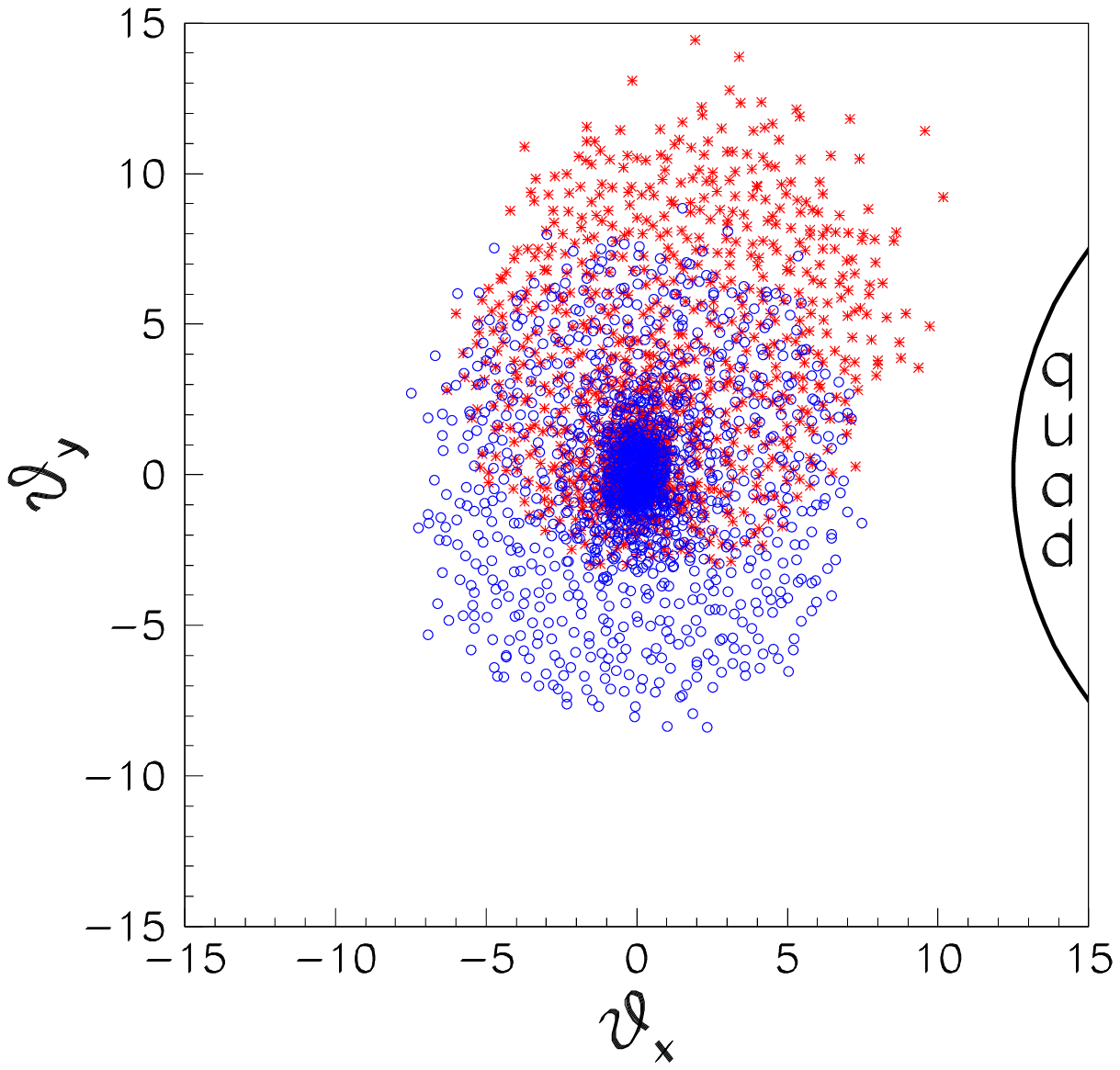}
\vspace{-0.7cm}
\caption{The shift of the outgoing beam due to the detector
  field. Blue (square) points: only beam-beam deflection, red (stars)
  points: the detector field of 4 T is added.  Positions of particles are
  taken at the distance of 4 m from the IP, at the place where they
  pass the first quad. Left figure: $2E_0=200$ GeV, right: $2E_0=500$
  GeV. }
\label{thxthy} \vspace{-0.0cm}
\end{figure}

Due to the crossing angle, the detector field gives an additional
deflection angle to the disrupted beam, see Fig.\ref{thxthy}.
A crab-crossing angle of 25 mrad is assumed. These figures correspond
to central collisions. For beams with an initial relative shift at the
IP, the central core is shifted due to the instability of collisions
but the maximum angles remain practically the same and decrease for
large beams shifts. One can see that particles get mostly a vertical
deflection, so the total vertical angle is about 17 mrad. The solenoid
field also leads to some horizontal displacement (due to the vertical
motion of particles) but it is smaller than the vertical shift of the
beam.

\begin{figure}[!htb]
\begin{minipage}{0.5\linewidth}
\hspace{0.cm}\includegraphics[width=6.cm, bb=0 0 563 550, clip]{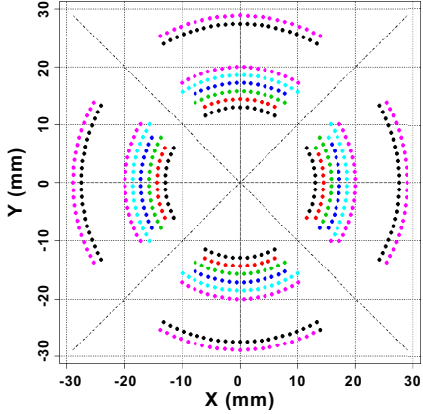}
 \vspace{-.2cm}
\end{minipage} \hspace{-0.cm}
\begin{minipage}{0.47\linewidth} \hspace{0mm}
 \vspace{-0.cm}
\caption{The design principle of a superconducting quad (only the coils are
  shown).  The radius of the quad with the cryostat is about 5 cm. The
  residual field outside the quad is negligibly small.}
\label{f-quad}
\end{minipage}
\vspace{-0.cm}
\end{figure}

A possible quad design for the photon collider was suggested by
B.~Parker \cite{Parker,TEL-Snow2005}, see Fig.~\ref{f-quad}.  The quad
consists of two quads of different radii, one inside another, with
opposite field directions. In this design, the gradient on the axis is
reduced only by 15\%, and the field outside the quad is practically zero.
The radius of the quad, the cryostat taken into account, is $R=5$ cm.
For the distance of the quad from the IP $L^*= $ 4 m and the
horizontal disruption angle of 12.5 mrad (10\% margin), the minimum
crab-crossing angle is $0.0125+5/400 = 25$ mrad. Obtaining the final
numbers requires some additional checks.

\begin{figure}[!htb]
\begin{minipage}{0.5\linewidth}
 \vspace{-0.cm}
\hspace{10mm}\includegraphics[width=6.cm,bb=0 0 482 475, clip]{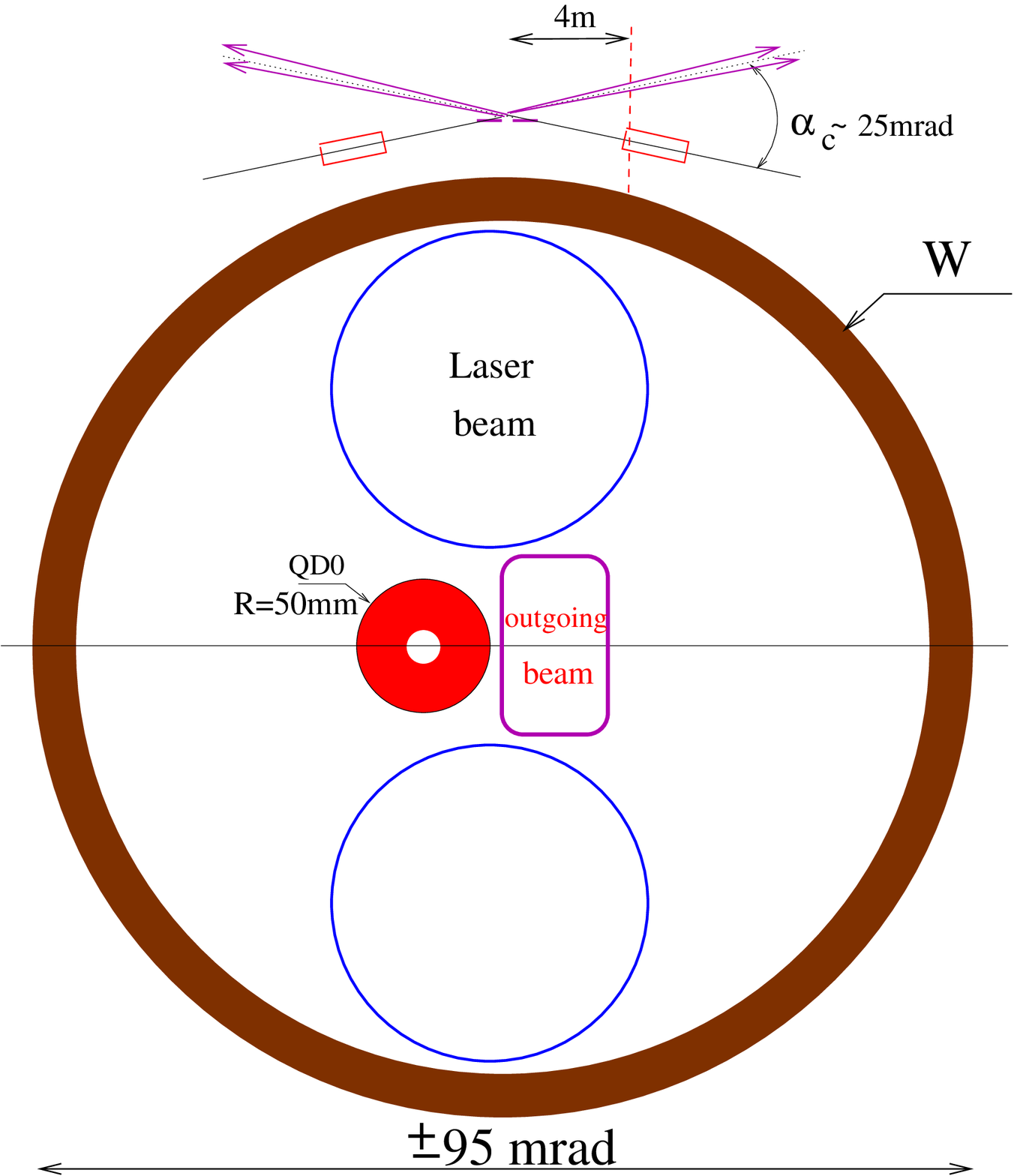}
\end{minipage} \hspace{-0.cm}
\begin{minipage}{0.47\linewidth}
\caption{Layout of the quad and electron and laser beams at the
  distance of 4 m from the interaction point (IP).}
\label{beams-quad}\end{minipage}
\end{figure}

Relative positions of the quad, the outgoing electron beam and the
laser beams at the distance 4 m from the IP is shown in
Fig.\ref{beams-quad}. We will return to this figure later when we
consider the laser optics.

\subsection{Other effects due to crossing angle}

{\bf Crab-crossing.}
In order to preserve the luminosity at large crossing angles, the
crab-crossing scheme is used, Fig.\ref{scheme}, where beams are tilted
by special RF crab-cavities. The requirements on the time and amplitude
stabilities of the RF become more stringent with the increase of the
crab-crossing angle. This problem is more important for the photon
collider where beams have smaller $\sigma_x$. For stabilization of the
crab-crossing angle, a fast feedback should be used based on the rate
of background processes (\EPEM\ pairs) and the azimuthal distributions of
outgoing particles (in the detector and beam dump), in addition to the
beam stabilization feedback based on beam deflection
(Sect.\ref{stab}). This problem need a detailed study. 

\begin{figure}[!htb]
\hspace*{2cm} \includegraphics[width=10.5cm]{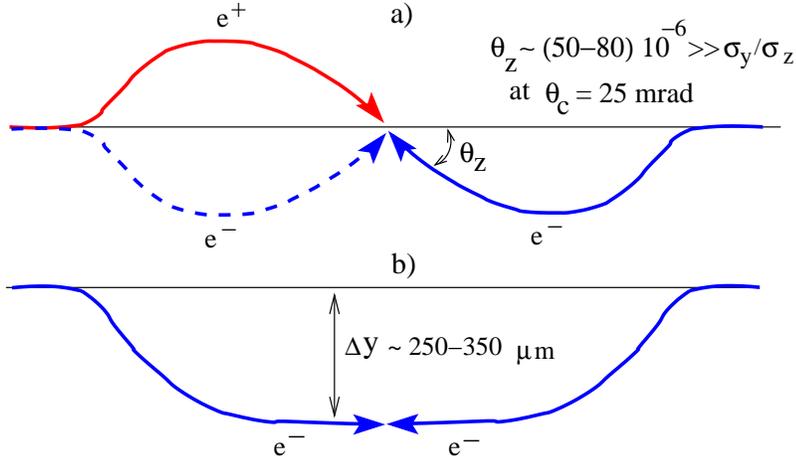}
\vspace{-0.4cm}
\caption{Trajectories of electrons (positrons) in the presence of the
solenoid field and a crab-crossing angle. At the lower figure, the
\EMEM\ collision angle is made zero using shifted quads.  }
\label{vert_angle}
\vspace{-0.0cm}
\end{figure}

{\bf Non-zero vertical collision angle.}
Due to the detector field, \EMEM\ beams collide at a non-zero vertical
collision angle that is several times larger than
$\sigma_y/\sigma_z$, Fig.\ref{vert_angle}. This angle can be removed
by a dipole correction winding in quads~\cite{telnov_lcws05_1113}.  Such
a correction shifts the IP vertically by about 300 \MKM, which is
an acceptable amount.

\begin{figure}[!htb]
\vspace{-0.5cm}
\begin{minipage}{0.5\linewidth}
\hspace{0.7cm}\includegraphics[width=7cm, height=7cm]{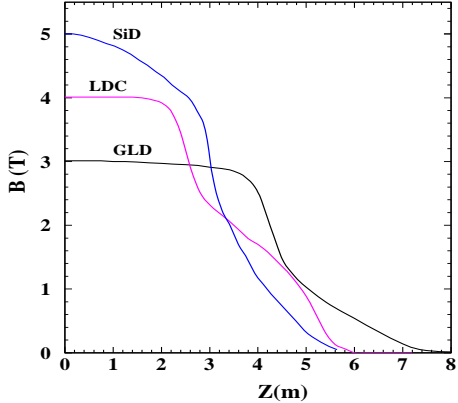}
\end{minipage}\hspace*{1.cm}
\begin{minipage}{0.4\linewidth}
\vspace{-1.5cm}
\caption{Magnetic field  $B(z,0,0)$ in LDC, SID and GLD detectors}
\label{bz}
\vspace{-0.cm}
\end{minipage}
\vspace{-1.cm}
\end{figure}

{\bf The increase of the vertical beam size due to SR.} Synchrotron
radiation (SR) in the detector field leads to an increase of the
vertical beam size. This effect was considered in
Refs.~\cite{telnov_lcws05_1113, TEL-Snow2005} for the detector fields
as of Summer 2005. In Spring 2006, the length of the LDC detector was
shorten from 7.4 m to 5.6 m. The simulation was repeated for
the detector field presented in Fig.\ref{bz}. Beam parameters
correspond to the nominal ILC case: $2E_0=500$ GeV, $N=2\times
10^{10}$, $\sigma_z=0.3$ mm, \ENX=$10\times 10^{-6}$~m,
\ENY=$0.04\times 10^{-6}$ m, $\beta_x=21$ mm, $\beta_y=0.4$ mm.

For the \GG\ case, instead of the \GG\ luminosity I simulated the \EMEM\
luminosity (without the $e\to\gamma$ conversion) with
$\sigma_y(\GG)=\sqrt{2}\sigma_y(\EPEM)$ in order to take into account an
effective increase of the vertical beam size due to Compton
scattering. All interactions between particles were switched off. The
position of the first quad (shifted in the \EMEM(\GG) case in order
to have a zero collision angle) was $z=3.8$--$6$ m for all detectors.
Results of the simulation are presented in Table~\ref{tab2}; the
statistical accuracy is about $\pm$0.5--1\%. 
{\small
\begin{table}[ht]
\bc
\caption{ Results on  $L(\alpha_c)/L(0)$.}  \setlength{\tabcolsep}{3.mm}
 {  \EPEM\ collisions} \\[2mm]
\begin{tabular}{lllllll}
$\alpha_c$(mrad) & 0 & 20 & 25 & 30 & 35 & 40 \\ \hline
LDC & 1. &0.997 &0.995 &0.99 &0.985 & 0.973 \\
SID & 1. &0.997 &0.993 &0.985 &0.97 & 0.93 \\
GLD & 1. &0.995 &0.99 &0.98 &0.96  & 0.935 \\
\end{tabular} \\[0.2cm]
{ \GG\ collisions} \\[2mm]
\begin{tabular}{lllllll}
$\alpha_c$(mrad) & 0 & 20 & 25 & 30 & 35 & 40 \\ \hline
LDC & 1 &0.996 &0.985 &0.963 &0.935 & 0.91 \\
SID & 1 &0.994 &0.99 &0.98 &0.955 & 0.93 \\
GLD & 1 &0.998 &0.993 &0.985 &0.973 & 0.94
\end{tabular}\\
\ec
\label{tab2}
\vspace{-.4cm}
\end{table}
} 
It is interesting that LDC is the best for \EPEM\ and the worst for \GG,
which is due to the compensation quadrupole in the \GG\ case. Its
contribution depends on the shape of the field at the location of the
quad (the radial field and the quad field have the same direction and therefore are added).
 
Conclusion: the crab-crossing angle of 25 mrad that is needed for the photon collider
is compatible with \EPEM.
\subsection{Beam dump}
\begin{figure}[!htb]
\vspace{-0.7cm}
 \hspace{1cm} \includegraphics[width=7cm]{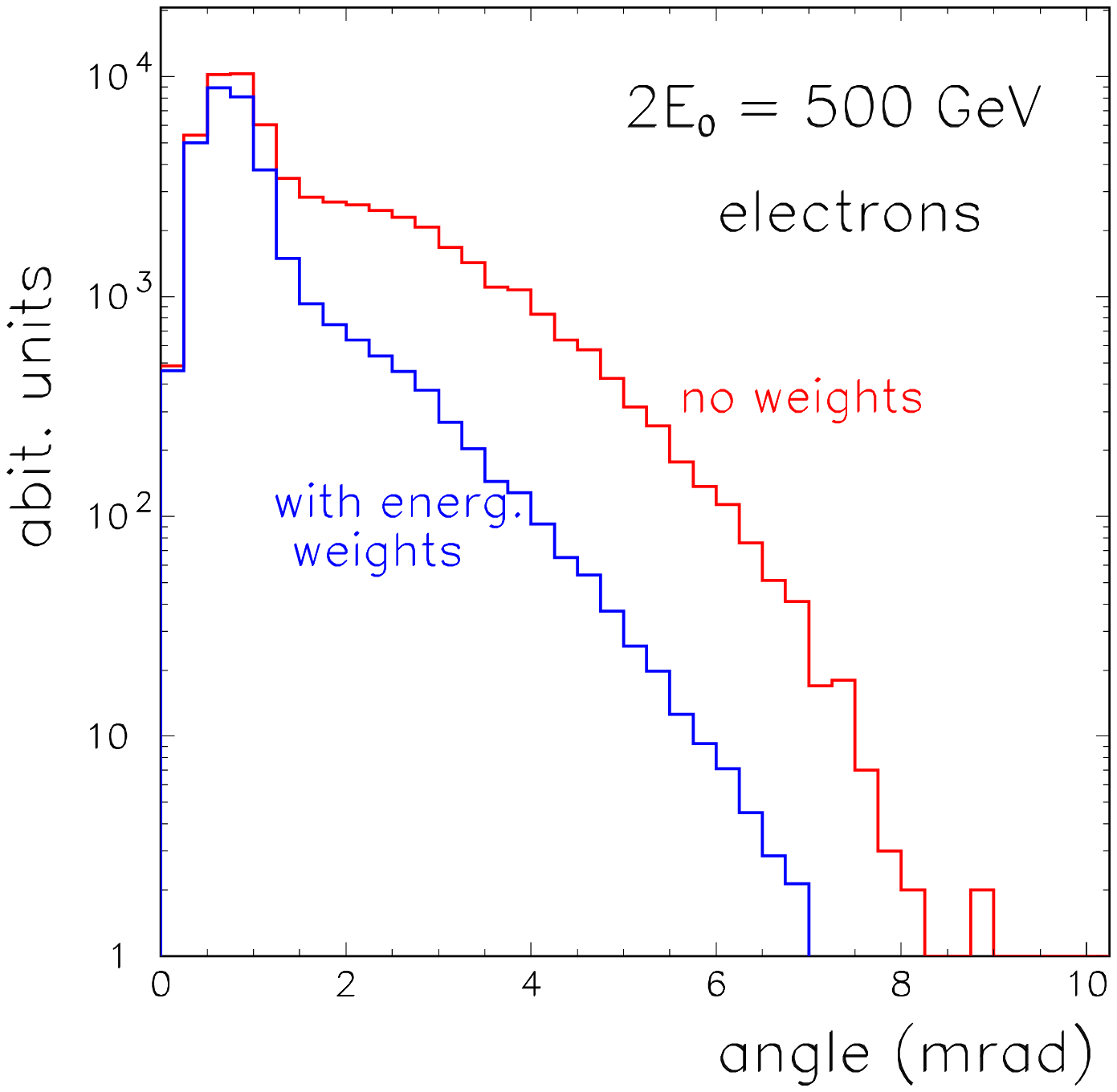}
\hspace{-0.cm}\includegraphics[width=7cm]{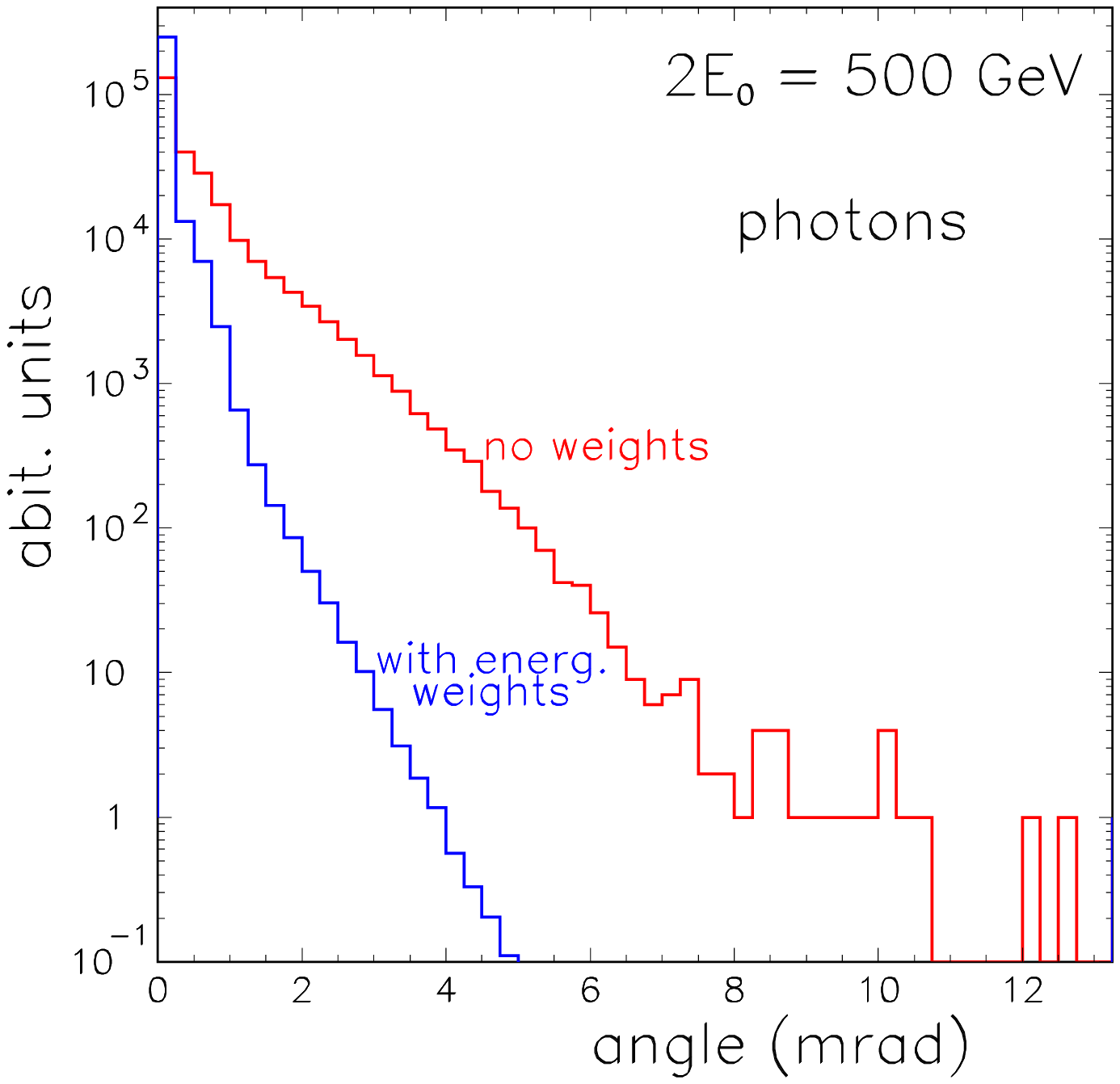}
\vspace{-0.5cm}
\caption{Angular distributions of electrons (left) and photons (right)
after the conversion and interaction points.}
\label{ang-dis}\vspace{-0.6cm}
\end{figure}

\begin{figure}[!htb]
\vspace{-0.3cm}
 \hspace{1.1cm} \includegraphics[width=7cm]{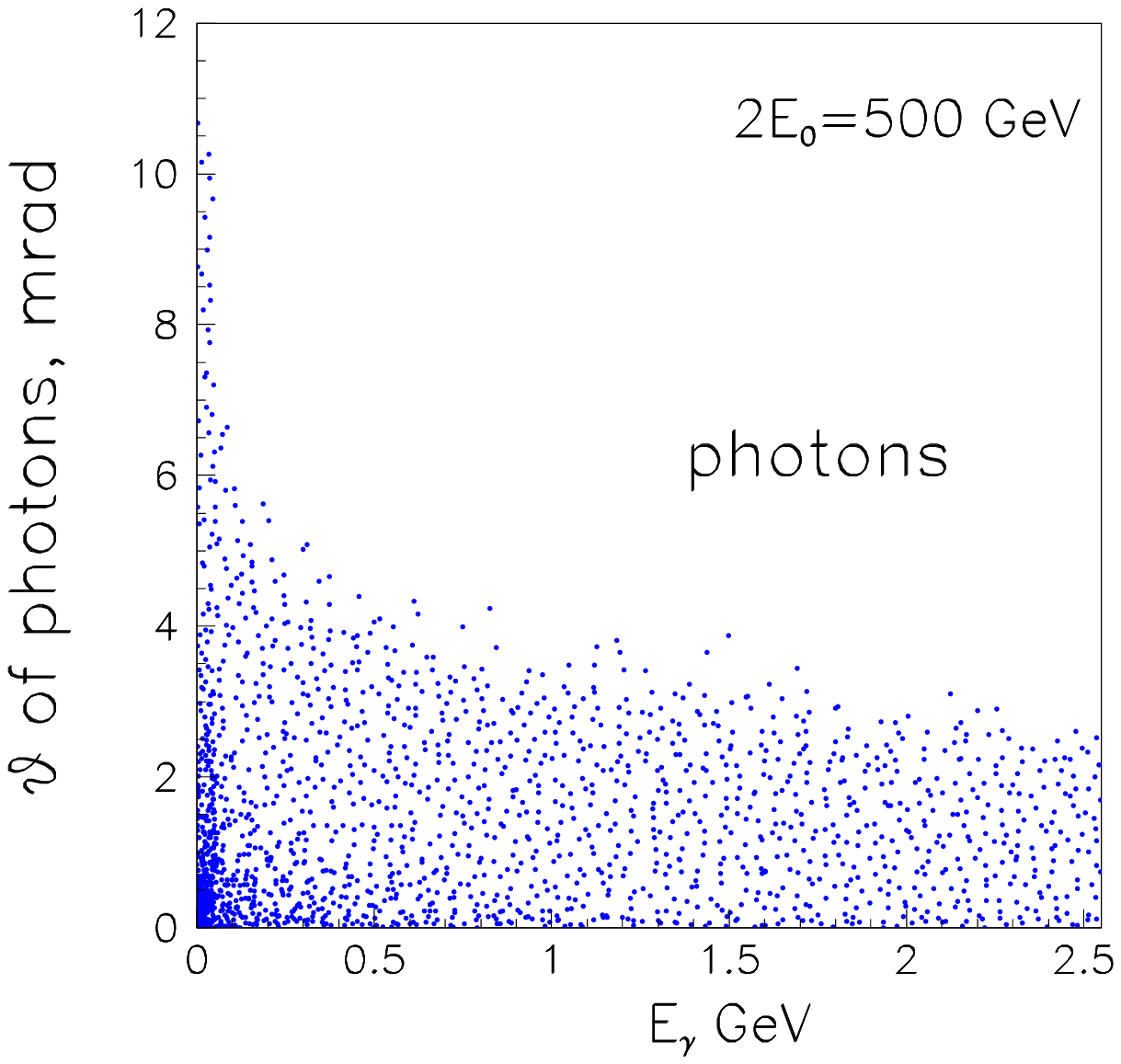}
\hspace{-0.6cm}\includegraphics[width=7cm]{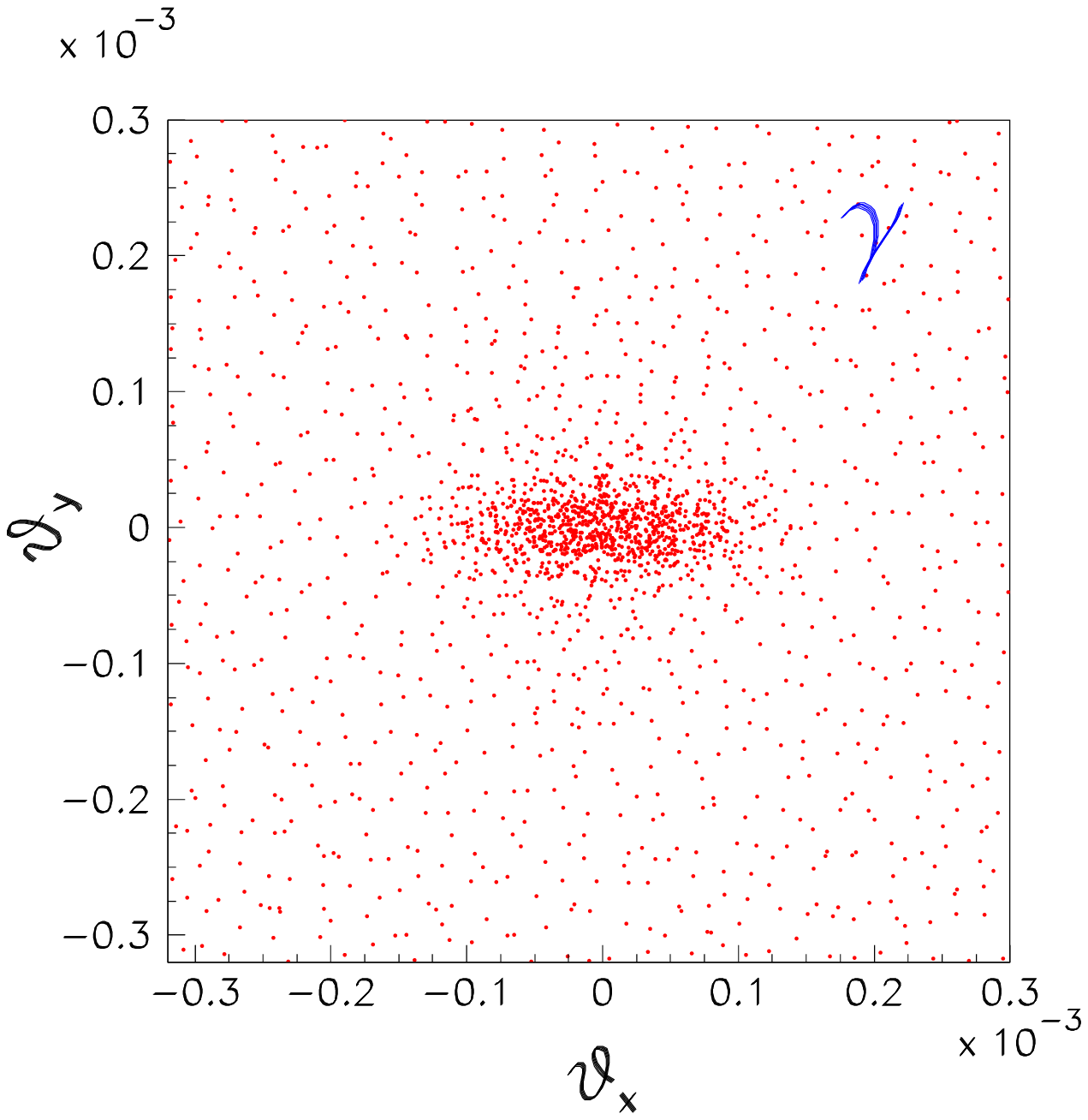}
\vspace{-0.9cm}
\caption{Energy-angular distributions of beamstrahlung photons (left) and the
angular distribution of Compton photons (right).}
\label{ang-p}
\vspace{-0.3cm}
\end{figure}

\begin{figure}[!htb]
 \centering
\includegraphics[width=14cm]{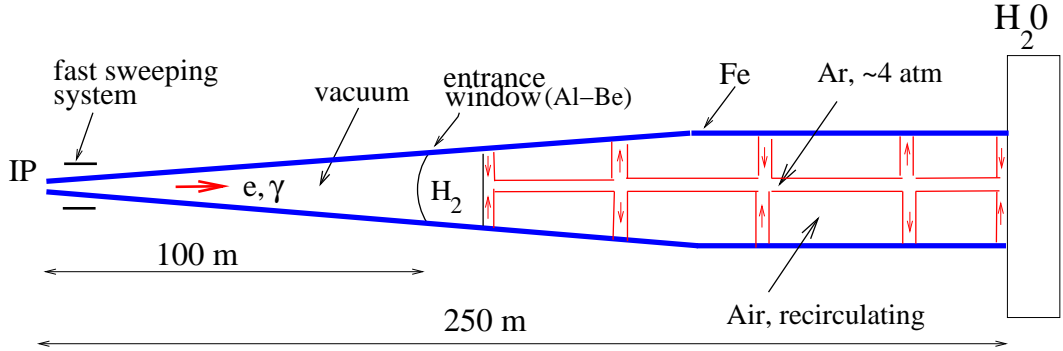}
\vspace{0.0cm}
\caption{An idea for the photon collider beam dump.  }
\vspace{-.3cm}
\label{beam-d}
\end{figure}

The photon collider needs a special beam dump, one that is very different from
the \EPEM\ beam dump. There are two main differences: \\[-4mm]
\bi
\item  The disrupted beams at a photon collider consist of an equal mixture of
  electrons and photons (and some admixture of positrons);\\[-4mm]
\item Disrupted beams at the photon collider are very wide (see 
  Fig.~\ref{ang-dis}), and need exit pipes of a large diameter.\\[-4mm]
\item On the other hand, the photon beam following the Compton scattering
  is very narrow. At the distance of 250 m from the IP, the r.m.s.
  transverse size of the photon beam is $1\times 0.35$ mm$^2$, see
  Fig.~\ref{ang-p}, with a power of about 10 MW. It cannot be dumped
  directly at a solid or liquid material. \\[-4mm]\ei
    There exists an idea of such a beam dump, as well as some
  simulations~\cite{Telnov-lcws04}, but a next step required, a  more
  careful study. The idea is the following. The water beam dump is
  situated at the distance of about 250 m from the IP, Fig.~\ref{beam-d}.
The electron beam can be swept by fast magnets (as in the TESLA
TDR) and its density at the beam dump will be acceptable. In order to
spread the photon beam we suggest placing a gas target, for example Ar
at $P\sim 4$ atm, at a distances of 120 to 250 m from the IP:
photons would produce showers, the beam diameter would increase, and
the density at the beam dump would become acceptable.  In order to
decrease the neutron flux in the detector, one can add a volume filled with
 hydrogen gas just before the Ar target, which would reduce the flux of
backward-scattered neutrons at the IP at least by one order of
magnitude. The corresponding numbers can be found in
Ref.~\cite{Telnov-lcws04}.

In order to reduce the diameter of the beamline between the beam dump
and the IP, it is desirable to slightly focus the disrupted electron
(positron) beam just after the exit from the detector (this issue has
not been considered yet). The angular distribution of beamstrahlung photons
is similar to that of beamstrahlung electrons that produced these
photons. However, the energy of beamstrahlung photons produced by the
rather  low-energy large-angle electrons is only a small fraction of
their energy, so the
effective (energy-weighted) angular distribution of photons is
narrower than that for electrons. According to Fig.~\ref{ang-dis}
(right), for photons a clearance angle of $\pm 3$ mrad will be sufficient,
which is 75 cm at the distance of 250 m.

  The Ar target should have a diameter of no more than 10 cm (a shower
  of such a diameter  does not present a problem for the beam dump).  The rest
  of the volume of the exit pipe with a diameter of about 1.5 m can be filled
   with air at 1 atm (or vacuum).  Such measures are necessary in
  order to avoid unnecessary scattering of low-energy electrons
  traveling at large distances from the axis and thus to
  reduce the energy losses and activation of materials (water, air) in the
  unshielded area (it is difficult to shield a 200 m tube).

\section{Configuration of the IP, transition from
  \EPEM\ to \GG}

  In order to save time and money, it is desirable to have an interaction region 
  that requires a minimum modification for transition from \EPEM\ to
  \GG\ collisions and back. The ideal case: the same beamlines and beamdumps, only
  minor modifications in the forward part of the detector. However,
  at present, the requirements presented by the \EPEM\ and \GG\ cases 
are very different and  no consensus reached. The differences are the  following:
\bi
\item \GG, \GE: the crab-crossing angle is 25 mrad (minimum), the
  outgoing beams go straight to the beam dump. Beams are very
  disrupted, so only the simplest diagnostics is possible, such as
  measurement of the beam profile in the beam dump area; 
  
\item \EPEM: the crab-crossing angle is 14--20 mrad, the extraction line
   includes many diagnostics such as precise measurement of
  the beam energy and polarization.  
 \ei
 
\begin{figure}[!htb]
\vspace{-0.1cm}
\bc \includegraphics[width=10cm]{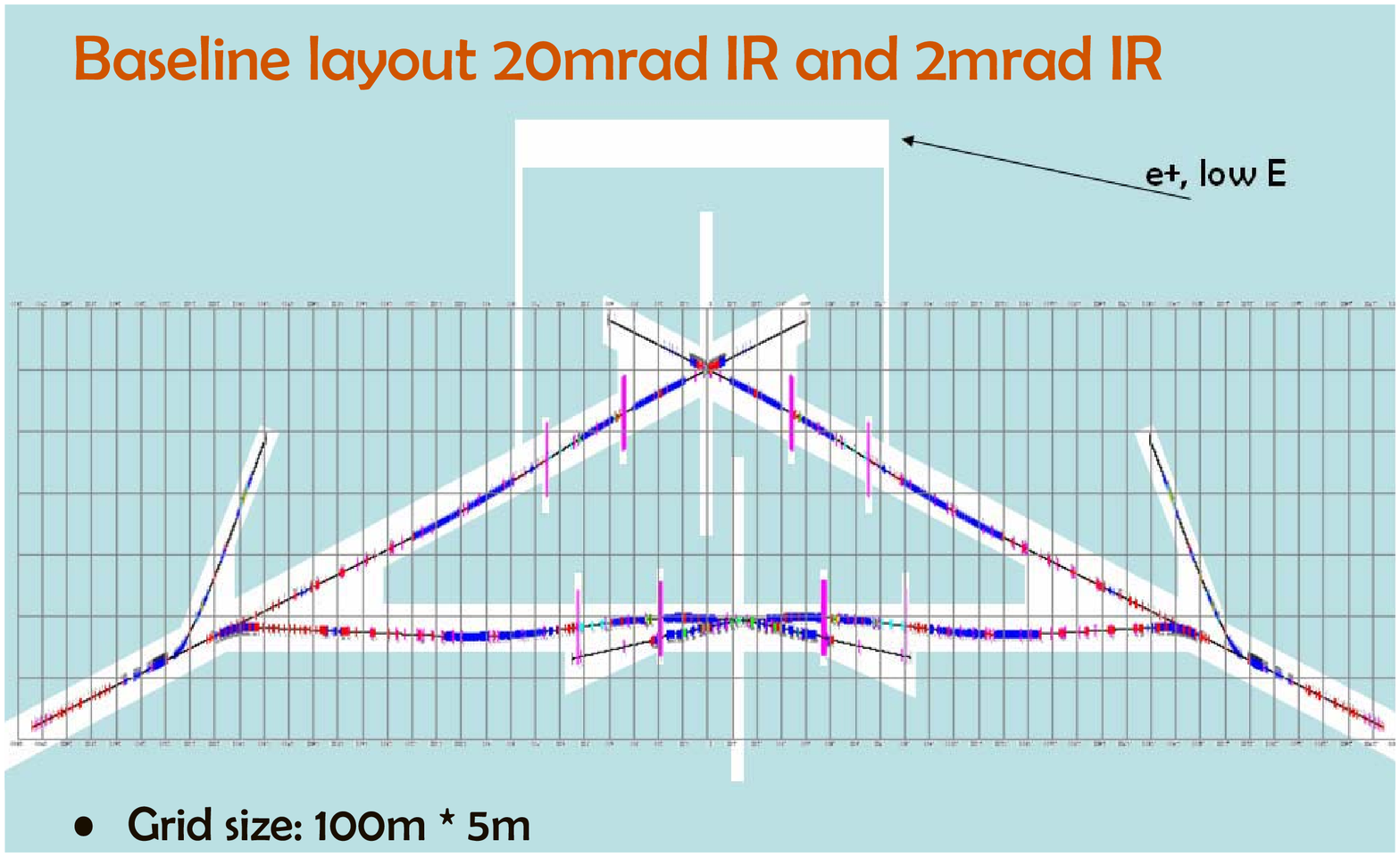}
\ec
\hspace{0.7cm}
\includegraphics[width=6.7cm]{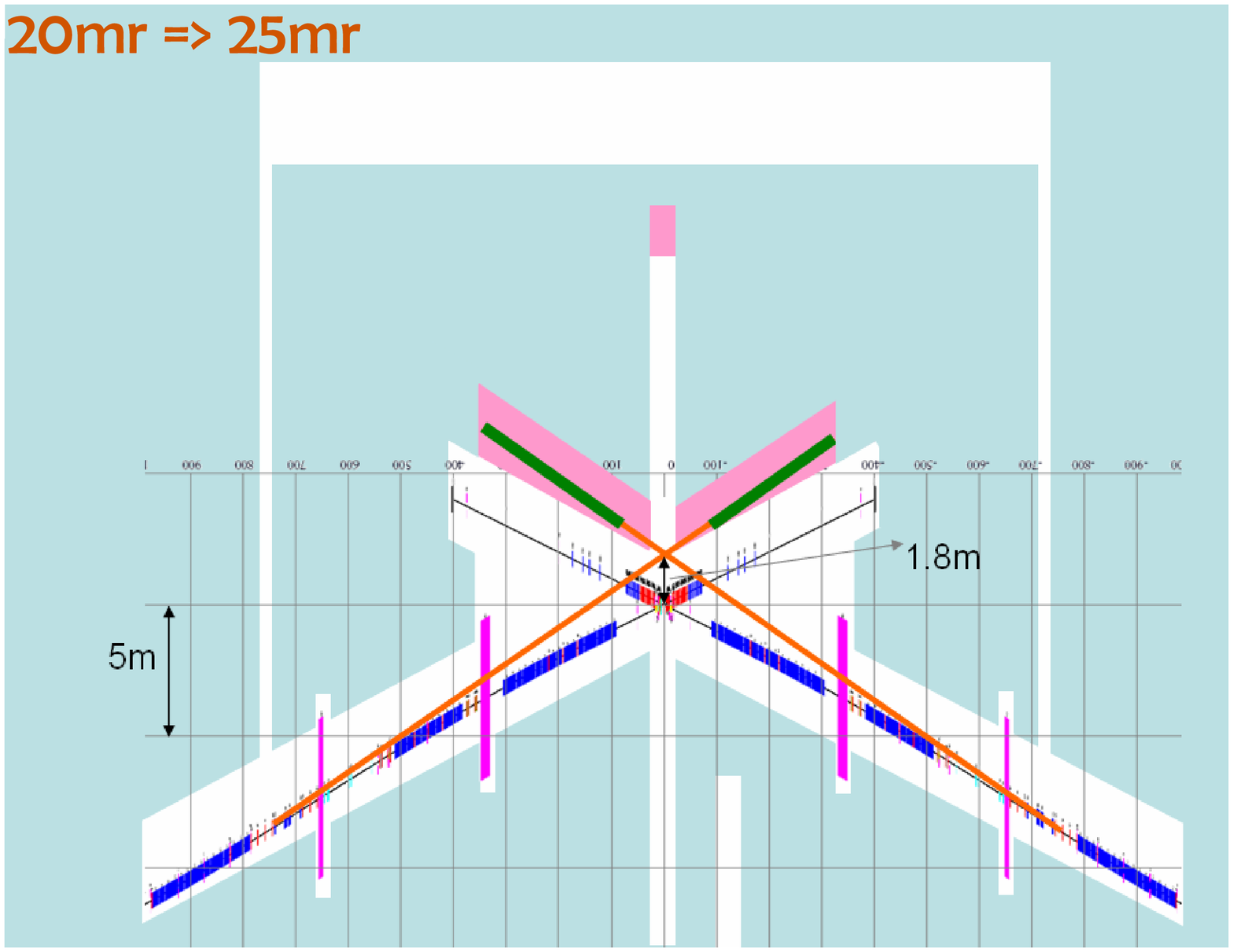}\hspace{0.2cm} \includegraphics[width=7.4cm]{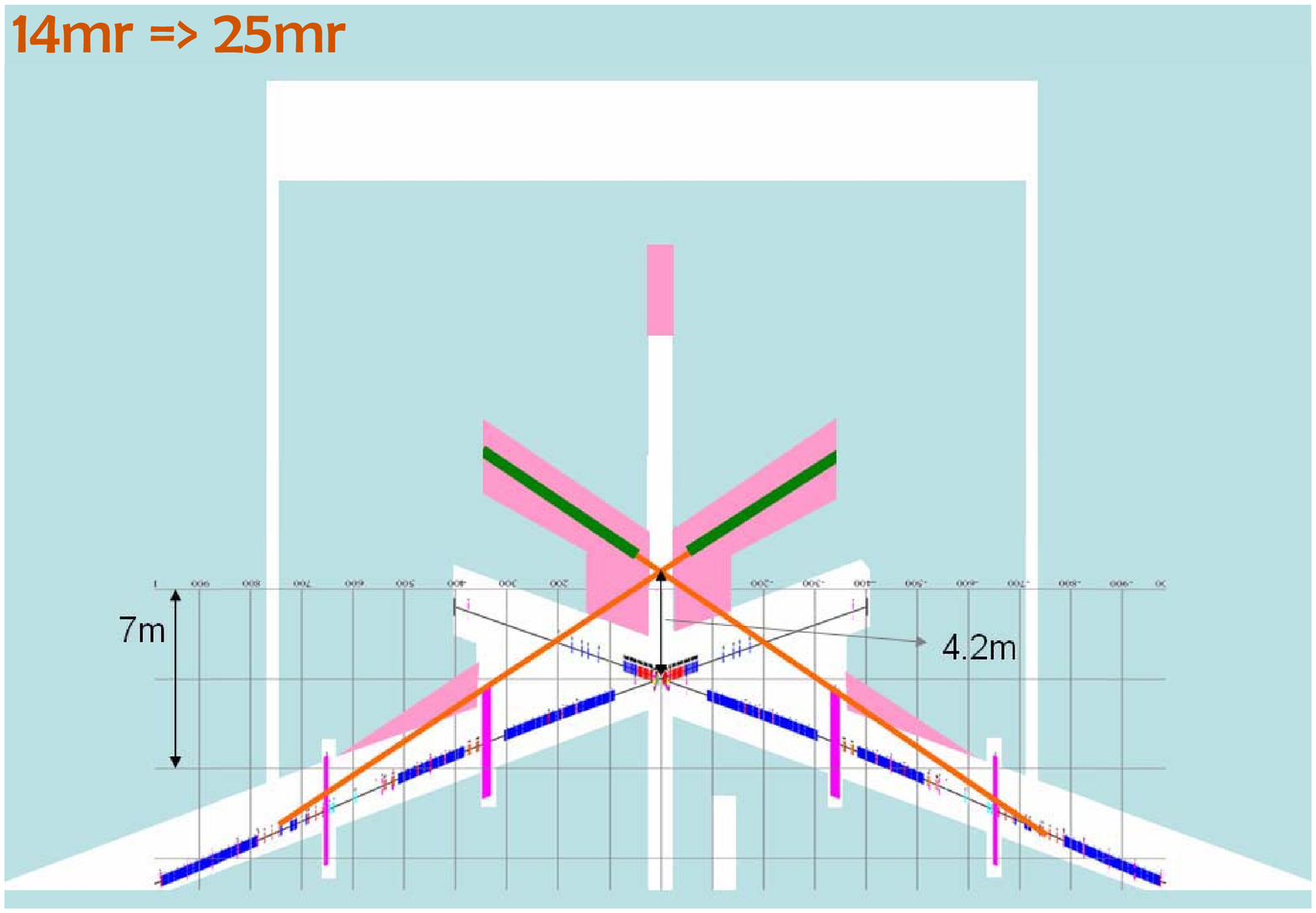}
\vspace{0.2cm}
\caption{Layout of the interaction regions at the ILC. The bottom figures
  show the upgrade path from \EPEM\ to \GG\ according to \cite{Seryi-lcws06}.
  See the author's alternative suggestion in the text below.}
\label{f:seryi}
\vspace{-0.05cm}
\end{figure}

 At present, the ILC beam delivery group has the following
 suggestion~\cite{Seryi-lcws06}. The extraction lines and the beam dump for \EPEM\ 
 and \GG\ are very different. Their replacements (transition to \GG\ 
 and back after the energy upgrade) will be problematic due to induced
 radioactivity. Therefore it makes sense to have different crossing
 angles and separate extraction lines and beam dumps for \EPEM\ and
 \GG. For the transition from \EPEM\ to \GG\ one has to move the detector
 and about 700 m of the up-stream beamline, Fig.\ref{f:seryi}. The
 displacement of the detector is equal to 1.8 m and 4.2 m for the increase of
 the crab-crossing angle from 20 to 25 mrad and from 14 to 25 mrad, respectively.

I have an alternative suggestion: the same crossing angle, the same
beam dump and no detector displacement. The cost will be reduced
considerably, by hundreds of millions of dollars! No time is needed for the shift of
beamlines (700 m!).  What are disadvantages? In this case, the designs
of the extraction line and the beamdump are dictated by \GG, so no
precision diagnostic in the extraction line for \EPEM\ is possible.
But is it really necessary?  Indeed, without such special extraction
line we can measure the energy and polarization before collisions,
many characteristics during the beam collision (the acollinearity angles,
distributions of the secondary \EPEM\ pairs, the beam deflection angles); we can
measure the angular distributions and the charged and neutral contents in the
disrupted beams. All this allows the reconstruction of the dynamics of beam
collisions, with a proper corrections in the simulation. For
example, the depolarization during the collision is rather small,
knowledge of beam parameters with a 10--20\% accuracy is sufficient for
introducing theoretical corrections. Direct measurement of the
polarization after the collisions does not exclude the necessity of such
a correction, it is just one additional cross check, but there are many
other cross checks besides the polarization.
 
 An additional remark. The requirement for the instrumented extraction line
 for \EPEM\ restricts the accessible set of beam parameters and
 correspondingly the luminosity. One can not use it for the case of
 large beamstrahlung losses. It will not work, for example, in the
 CLIC environment or at the photon collider. In other words, such
 diagnostic of outgoing beams is useful but not absolutely necessary at linear
 colliders.
 
 This suggestion is very attractive, cost- and time-effective,
 and deserves a serious consideration by appropriate GDE committees.
   
\section{The laser system}
   
 The laser parameters required for the photon collider:\\[-6mm]
\bi
\item Wavelength \hspace*{0.2cm} $\sim 1$ \MKM\  (good for $2E_0 < 700$
  GeV); \\[-7mm]
\item Time structure \hspace*{0.cm} $c \Delta t \sim 100$ m, 3000 bunches/train; \\[-7mm]
\item Flash energy \hspace*{0.2cm}$\sim 9$ J  (about one scattering length for
  $E_0=250$ GeV); \\[-7mm]
\item Pulse length  \hspace*{0.4cm} $\sigma_t \sim 1.5$ ps.\\[-7mm]
\ei

The most attractive scheme for a photon collider with the ILC pulse
structure is storage and recirculation of a very powerful laser pulse
in an external optical cavity~\cite{Tfrei,
  TEL2001,Will2001,TESLATDR,Klemz2005}. This can reduce the required
laser power by a factor of $Q \sim 100$ ($Q$ is the quality factor of
the cavity).

\begin{figure}[!htb]
\begin{minipage}{0.5\linewidth}
\vspace{-0.7cm} \hspace{-0.5cm}
\includegraphics[width=6.5cm]{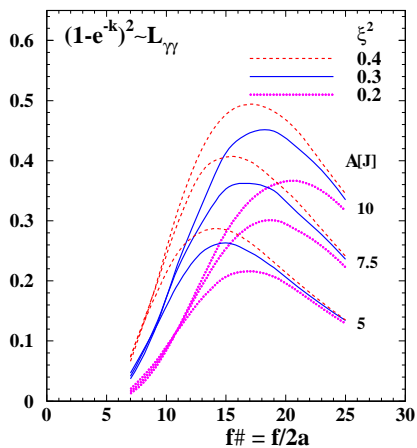}
\end{minipage}
\begin{minipage}{0.45\linewidth}
\vspace{-1.5cm}
\caption{Dependence of  \LGG\ on the flash energy and $f_{\#}$
(flat-top laser beam) for several values of the parameter $\xi^2$. }
\label{conversion}
\end{minipage}
\vspace{-1.3cm}
\end{figure}

Dependence of the \GG\ luminosity on the flash energy and
$f_{\#}=F/2R$ (flat-top laser beam) for several values of the
parameter $\xi^2$ (which characterizes the multi-photon effects in Compton
scattering, $\xi^2 < 0.3$ is acceptable \cite{TESLATDR}) is presented
in Fig.~\ref{conversion}~\cite{TEL-Snow2005}.  This simulation is
based on the formula for the field distribution near the laser focus
for flat-top laser beams. It was assumed that $\alpha_c = 25$ mrad and
the angle between the horizontal plain and the edge of the laser beam
is 17 mrad (the space required for disrupted beams and quads, see
Fig.~\ref{beams-quad}). At the optimum, $f_{\#} \sim 17$, or the
angular size of the laser system is about $\pm 0.5/f_{\#} \approx \pm
30$ mrad.  If the focusing mirror is situated outside the detector at
the distance of 15 m from the IP, it should have a diameter of about 1
m.  All other mirrors in the ring cavity can have smaller diameters,
about 20 cm is sufficient from the damage point-of-view (diffraction
losses require an additional check).

The DESY--Zeuthen group has  considered an optical cavity
at the wave level, its pumping by short laser pulses, diffraction
losses, etc.~\cite{Klemz2005}. 

In the design with the final mirror located outside the detector, at a
distance $\sim 15$ m from the detector center, the mirror's diameter is very
large, about $d\sim 90$ cm, and the open angle in the detector as
large as $\pm 95$ mrad is required. The detectors that are currently under consideration for
\EPEM\ have holes in the forward directions of about $\pm$ 33 -- 50
mrad. Modifying the for \GG\ required that parts of ECAL, HCAL
and the yoke be removable.

\begin{figure}[!htb]
\centering
\includegraphics[width=10cm]{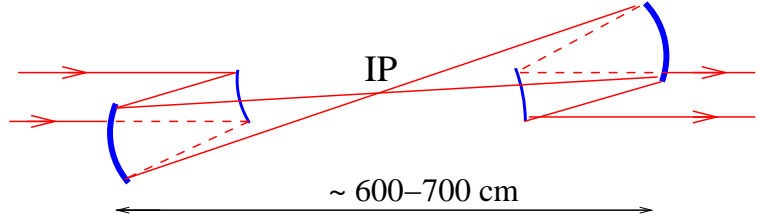}
\caption{Laser optics inside the detector (alternative scheme). }
\label{cavity}
\vspace{-0.0cm}
\end{figure}

An alternative scheme was considered in the TESLA TDR: the final pairs of
mirrors are situated inside the detector, Fig.\ref{cavity}. In this case, the
diameter of the focusing mirrors is only 20 cm and that of auxiliary
mirrors is about 11 cm. The dead angle for tracking remains, as before,
about $\pm 95$ mrad; it can be smaller for the calorimeters, and may be the
same as for \EPEM. The laser density at the mirrors is far from the
damage threshold, the average power is the most serious
problem~\cite{TESLATDR}.

Though the cavity reduces substantially the required laser energy, the
laser should still be very powerful. According to a LLNL estimation, the cost of
one such laser is about \$ 10 M \cite{Gronberg-snow05}.  The photon
collider needs two such lasers and one or two spares.

\begin{figure}[thb]
\vspace{-2.3cm}
\hspace{0.cm}\includegraphics[width=13.9cm]{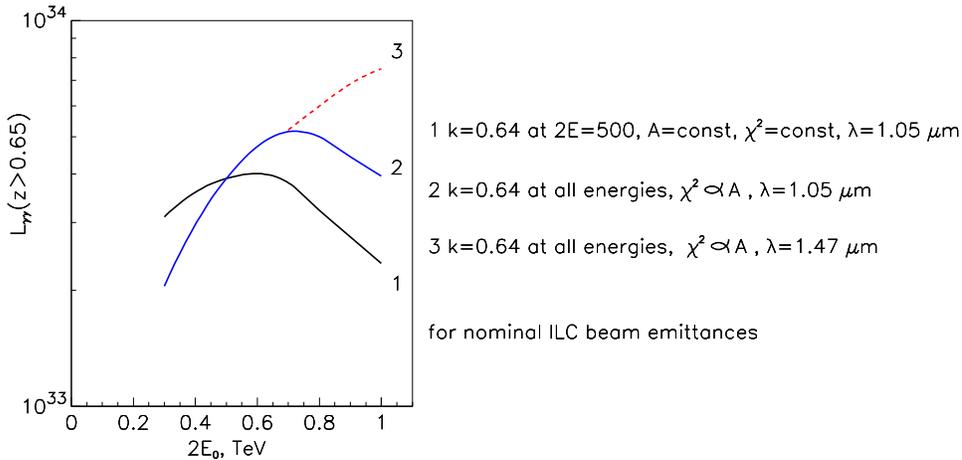}
\vspace{-1.9cm}
\caption{Dependence of the \GG\ luminosity on the energy.}
\label{lgg} \vspace{-0.0cm}
\end{figure}

The same laser with the 1 \MKM\ wavelength can be used up to the ILC
energy $2E_0 \sim 700$ GeV. At higher energies, the \GG\ luminosity
decreases due to \EPEM\ pair creation in the conversion region in
collision of the high-energy and laser photons~\cite{GKST83,TEL95} and
due to the decrease of the Compton cross section, see
Fig.~\ref{lgg}~\cite{TEL-Snow2005}.  For the energy $2E_0=1$ TeV, the
reduction in the luminosity due to this effect is about a factor of
2--3 compared to the optimum case.  For the high energies it is
desirable to have a wavelength of about 1.5--2 \MKM. The technical
feasibility of such a laser has not been studied yet.

\section{Conclusion}

In summary I would like to stress several important issues that need 
urgent attention of the ILC designers (GDE).
\begin{itemize}

\item We need a clear and inexpensive path from the \EPEM\ to \GG, \GE\ 
  modes of operation. The best would be an IP with a crab-crossing
  angle of about 25 mrad both for \EPEM\ and \GG. This is possible, but
  an effort is required to reach a consensus in the physics community.  The presently
  suggested upgrade pass is too difficult, considerably increases the
  ILC cost and is time-consuming.

\item In order to achieve high luminosity at the photon collider,
damping rings with emittances that are much smaller than for \EPEM\ are required. 
A serious and detailed study of this problem is needed. It is not excluded that
a optimized wiggler-dominated storage ring will allow a x3 -- x5 higher
luminosity than that in the present design.
  
\item The photon collider is not ``an option'' that can be
  implemented some time later --- it is an integral part of the ILC project and
  considerably influences the baseline designs of practically all ILC
  systems. In addition, the key element of the project is a very
  unique, state-of-the-art laser system whose development required substantial time and
  finances.  The photon collider can be
  successfully built only if it is an integral part of the \EPEM,
  \GG, \GE, \EMEM\ linear collider.
\end{itemize}

 \section*{Acknowledgements}
 I would like to thank Maria Krawczyk for her great efforts on organization of
 PHOTON2005 in Warsaw and PLC2005 in Kazimierz and
 creating a beautiful and friendly atmosphere at the conferences.

\end{document}